\documentclass[12pt,a4paper]{article}
\usepackage[utf8]{inputenc}
\usepackage{natbib}

\usepackage[margin=1in]{geometry}

\usepackage{physics}
\usepackage{amsthm, amssymb, amsmath, appendix, mathtools}
\usepackage{hyperref}
\usepackage{enumitem}
\usepackage{cancel}  
\usepackage{setspace}

\mathtoolsset{showonlyrefs=true}

\usepackage[mathlines]{lineno}
\usepackage{tikz}
\usetikzlibrary{intersections, calc, arrows.meta}
\usepackage{cancel}
\usepackage{subcaption}

\makeatletter

\newtheoremstyle{allcaps_style}
  {\dimexpr\topsep + 1mm\relax}  
  {\dimexpr\topsep + 1mm\relax}  
  {\itshape}                     
  {}                             
  {\normalfont}                  
  {:}                            
  {0.5em}                        
  {\MakeUppercase{\thmname{#1}}\thmnumber{ #2}\thmnote{ (#3)}}

\newtheoremstyle{allcaps_style_def}
  {\dimexpr\topsep + 1mm\relax}  
  {\dimexpr\topsep + 1mm\relax}  
  {\normalfont}                  
  {}                             
  {\normalfont}                  
  {:}                            
  {0.5em}                       
  {\MakeUppercase{\thmname{#1}}\thmnumber{ #2}\thmnote{ (#3)}}

\theoremstyle{allcaps_style_def}

\theoremstyle{allcaps_style}
\newtheorem{theorem}{Theorem}

\newtheorem{lem}{Lemma}
\newtheorem{proposition}{Proposition}

\newtheorem{axiom}{Axiom}

\renewcommand{\proofname}{PROOF:}



\usepackage{titlesec}

\titleformat{\section}
  {\normalsize\bfseries\centering} 
  {\thesection.} 
  {0.5em} 
  {} 

\titleformat{\subsection}
  {\normalfont\centering} 
  {\thesubsection.}    
  {0.5em}              
  {\itshape}    

\titleformat{\subsubsection}
  {\normalfont\centering} 
  {\thesubsubsection}    
  {0.5em}              
  {}    

\titleformat{name=\section, numberless}
  {\normalsize\centering\scshape}
  {}
  {0em}
  {}

\newcommand{\ssd}{\trianglerighteq_\text{D}}
\newcommand{\F}{\mathcal{F}}
\newcommand{\cP}{\mathcal{P}}
\newcommand{\bP}{\mathbb{P}}
\newcommand{\cQ}{\mathcal{Q}}
\newcommand{\bQ}{\mathbb{Q}}
\newcommand{\bR}{\mathbb{R}}

\newcommand{\nrsuccsim}{\, \cancel{\succsim^+} \,}

\title{Partially rational preferences under ambiguity\thanks{The authors are grateful to Kazuhiro Hara, Kaname Miyagishima, Satoshi Nakada, Hendrik Rommeswinkel, Koichi Tadenuma, and Norio Takeoka for their helpful comments and insightful discussions. This paper was presented at a seminar at Waseda University.}}
\author{Kensei Nakamura\thanks{Graduate School of Economics, Hitotsubashi University, Kunitachi, Tokyo 186-8601, Japan. E-mail: kensei.nakamura.econ@gmail.com}\hspace{1mm} and Shohei Yanagita\thanks{Graduate School of Economics, Hitotsubashi University, Kunitachi, Tokyo 186-8601, Japan. E-mail: shoheiyanagita@gmail.com}}
\date{\today}

\begin{document}

\onehalfspacing
\sloppy

\maketitle


\begin{abstract}
    Completeness and transitivity are standard rationality conditions in economics. However, under ambiguity, decision makers sometimes violate these requirements because of the difficulty of forming accurate predictions about ambiguous events. 
    Motivated by this, we study various ambiguity preferences that partially satisfy completeness and transitivity. 
    Our characterization results show that completeness and a novel yet natural weakening of transitivity correspond to two opposite ways of using multiple probability distributions in mind; that is, 
    these two axioms have dual implications at the level of cognitive processes for ambiguity.
    \vspace{3mm}
\end{abstract}

\vspace{3mm}



\section{Introduction}
\label{sec:intro}

Ambiguity refers to situations in which the true probability law over states is either unknown or nonexistent, 
and is inherently involved in many decision problems studied in economics.
In the standard decision-making model known as the subjective expected utility (SEU) model, decision makers (DMs) have a unique probability distribution as their belief and evaluate each prospect according to its expected utility level. 
However, because of the lack of objective information about the likelihood of states, 
real-world DMs sometimes fail to form a single prior as their belief. 
As a result, they cannot always behave in the organized manner presumed in the SEU model. 

Rather, DMs facing ambiguity have multiple probability distributions in their minds and compare ambiguous prospects using them. 
Such multiplicity of beliefs has been extensively studied in decision theory; moreover, it is well known to be closely related to violations of rationality conditions---completeness and transitivity---assumed in the SEU model.

Indeed, \citeauthor{bewley2002knightian}'s (\citeyear{bewley2002knightian}) seminal paper showed that violations of completeness can explain the multiplicity of probability distributions in DMs' minds. 
According to Bewley's model, the DM evaluates one prospect $f$ to be weakly better than another one $g$ if the expected utility level of $f$ is not less than that of $g$ for every probability distribution in the DM's mind. 
Also, \citet{lehrer2011justifiable} examined the opposite of Bewley's model---the multiprior model in which the DM evaluates one prospect $f$ to be weakly better than another one $g$ if there is at least one probability distribution in mind such that the expected utility level of $f$ is weakly higher than that of $g$. 
These preferences are called justifiable preferences. 
\citeauthor{lehrer2011justifiable} showed that the key difference from the standard SEU model can be explained by violations of transitivity. 
An important observation from these results is that the ways of processing multiple beliefs are closely related to how the rationality conditions are violated.

Compared to the above representative models, real-world DMs exhibit varying degrees of rationality. 
Examining how these deviations from the standard model are related to cognitive processes for ambiguity can shed light on the role of rationality axioms widely assumed in economics.  
Motivated by this, the present paper investigates partially rational preferences, focusing on two rationality axioms: completeness and a weakening of transitivity.
Completeness is the standard axiom that ensures that the DM can compare any pair of alternatives. Our weakened version of transitivity requires that if a constant prospect $x$ is indirectly better than another constant $y$ via the relation with an arbitrary prospect $f$, then $x$ is better than $y$. 
Thus, this axiom ensures that the constant prospects serve as the basis for evaluating ambiguous prospects by prohibiting preference reversals of constant ones around an arbitrary alternative. 
Both are implicitly assumed conditions in all preferences under ambiguity represented by utility functions.

We study their implications within a general class of preferences that includes both the Bewley and justifiable models as special cases. 
A preference in this class can be represented by a pair $(u, \mathbb{P})$ of a von Neumann-Morgenstern (vNM) function and a collection of sets of beliefs. 
This model can be easily understood by considering a DM with multiple selves in mind. 
Suppose that each self has a set of priors $\cP\in \bP$ and makes suggestions to the DM based on the Bewley model. 
According to this model, the DM evaluates one prospect to be weakly better than another if at least one self reaches that conclusion. 
We call these preferences generalized Bewley preferences. 
It is known that these preferences can be characterized by weakening completeness and transitivity in a certain way from the SEU model (\citet{lehrer2011justifiable}, \citet{nakamura2025cautiousdualselfexpectedutility}). 
Furthermore, it should be noted that for the generalized Bewley preference associated with $(u, \mathbb{P})$, we can represent the same preference using another pair $(u, \mathbb{Q})$ as the unanimity rule among the selves with justifiable preferences. 
Thus, the important point is that, in both representations, these preferences can be written as combinations of the for-all and for-some rules, and the ordering of these rules is not important at the preference level.

Our main characterizations encapsulate the implications of completeness and our weak transitivity as cognitive procedures of multiple priors. 
In these results, the opposite effects of completeness and (our weak) transitivity play a key role. 
Under completeness, preferences need to be sufficiently decisive, so this axiom suggests that preferences reach weak preference relations easily enough. 
On the other hand, transitivity can be understood as requiring that the DM cannot reach a weak relation if it yields some inconsistent choice pattern. 
Thus, under this condition, the DM should be humble enough. 
The conceptual contribution of this paper is to find that these polar opposite effects can be represented in a unified way through cognitive processes. 


The first theorem shows that the generalized Bewley preferences that satisfy completeness can be characterized by a new way of processing multiple priors. 
These preferences are again parameterized by a pair $(u, \mathbb{P})$. 
According to this model, the DM first compares a pair of alternatives by the for-some comparison of Bewley judgments and the for-all comparison of justifiable judgments. 
The DM then concludes that a prospect is weakly better than another if at least one of these comparisons supports that evaluation. 
Thus, our results finds that considering the \textit{disjunction} of the two dual processes can ensure the decisiveness that follows from completeness. 

The second theorem shows that our weak transitivity has the opposite implication of completeness at the representation level: 
The generalized Bewley preferences with the weakened version of transitivity can be represented as the \textit{conjunction} of the two dual processes. 
That is, the DM concludes that one prospect is weakly better than another if both of the for-some comparison of Bewley judgments and the for-all comparison of justifiable judgments support that evaluation. 
By taking the intersection of the two comparisons, the DM becomes less likely to conclude that one prospect is better than another, thereby preventing inconsistent choice patterns prohibited by transitivity.
Our theorem formally establishes the connection between this procedure and the weak transitivity. 

Finally, we examine the joint implication of these two properties. 
We show that comparisons by a generalized Bewley preference with completeness and the weakened version of transitivity can be represented by the \textit{average} of the relative benefits of the two dual comparisons.  
According to our theorem, this averaging can balance the opposite effects on the degree of decisiveness and thereby characterize their joint implications.  

To summarize, we find that the rationality axioms are closely related to how DMs process multiple priors.
Since the axioms studied in this paper are normatively appealing, our representation theorems offer new insights into collective decision-making problems in which each agent has multiple priors (e.g., \citet{danan2016robust}).
In the collective decision context, each prior set can be considered as an expert's belief, and our theorems can be viewed as suggesting how the collective DM should aggregate the experts' opinions. 

This paper is organized as follows.
Section \ref{sec:pre} introduces the formal setup and the generalized Bewley preferences. 
Section \ref{sec:main} examines the implications of completeness and a weakening of transitivity. 
These axioms have comprehensive geometric implications, so we also discuss the graphical intuition behind the theorems. 
Section \ref{sec:alt} provides supplementary results---parametric characterizations of the two axioms and characterizations of counterpart rationality axioms for the negative part. 
Finally, Section \ref{sec:liter} discusses the related literature.
All proofs are in Appendix.

\section{Preliminary}
\label{sec:pre}

\subsection{Framework}

We study \citeauthor{anscombe1963definition}'s (\citeyear{anscombe1963definition}) framework elaborated by \cite{fishburn1970book}. 
Let $S$ be a finite set of states.  
There is a set of deterministic prizes such as monetary payoffs, and let $X$ be the set of probability measures with finite support over the prize set.
An \textit{act} is a mapping $f: S \rightarrow X$, and the set of acts is denoted by $\mathcal{F}$.  
As usual, we identify a lottery $x\in X$ with the constant act $f$ such that $f(s) = x$ for all $s \in S$. 
For $f,g \in \mathcal{F}$ and $\alpha \in [0,1]$, let $\alpha f + (1 - \alpha) g$ be the act such that for all $s\in S$, $(\alpha f + (1 - \alpha) g)(s) =  \alpha f(s) +(1- \alpha) g(s)$. 

A DM has a binary relation $\succsim$ over $\mathcal{F}$. 
As usual, for $f, g\in \mathcal{F}$, $f \succsim g$ means that the DM weakly prefers $f$ to $g$. 
The asymmetric and symmetric parts of $\succsim$ are denoted by $\succ$ and $\sim$, respectively. 

Let $\Delta (S)$ be the set of probability distributions over $S$. We embed $\Delta(S)$ in $\mathbb{R}^S$ and endow $\Delta (S)$ with the Euclidean topology. 
Its typical element is denoted by $p$ or $q$. 
With a slight abuse of notation, for $f\in \mathcal{F}$ and $u: X \to \mathbb{R}$, let $u(f) \in\mathbb{R}^S$ be such that $u(f) (s) = u(f(s))$ for all $s\in S$. 
For $p\in \Delta(S)$ and $\varphi\in \mathbb{R}^S$, let $\mathbb{E}_p [\varphi] = \sum_{s\in S} p(s) \varphi (s)$. 


We say that a binary relation $\succsim$ is an \textit{SEU preference} if there exist a nonconstant affine function $u:X\to \mathbb{R}$ and $p\in \Delta(S)$ such that for all $f,g\in \mathcal{F}$, 
\begin{equation*}
    f\succsim g \iff \mathbb{E}_p [u(f)] \geq \mathbb{E}_p [u(g)]. 
\end{equation*}
\subsection{Basic Axioms and a Benchmark Result}

The following two rationality axioms have been assumed in most economic analyses. 

\begin{axiom}[Completeness]
    For all $f,g\in \mathcal{F}$, $f\succsim g$ or $g\succsim f$.  
\end{axiom}

\begin{axiom}[Transitivity]
    For all $f,g,h\in \mathcal{F}$, if $f\succsim g$ and $g\succsim h$, then $f\succsim h$. 
\end{axiom}

However, as discussed in Section \ref{sec:intro}, DMs in the real world sometimes violate these axioms because of the lack of objective information. 
In the following, we introduce a baseline decision-making model that can account for these deviations (\citet{lehrer2011justifiable,nakamura2025cautiousdualselfexpectedutility}). 

We begin with the minimal requirements for binary relations over $\mathcal{F}$.

\begin{axiom}[Non-Triviality]
\label{axm:nontri}
    For some $f,g\in \mathcal{F}$, $f\succ g$. 
\end{axiom}

\begin{axiom}[Reflexivity]
    For all $f\in \mathcal{F}$, $f\succsim f$. 
\end{axiom}

We consider preferences that can violate \textit{completeness} and \textit{transitivity}, but we impose minimal conditions on rationality to focus on preferences where violations of rationality only come from ambiguity, not from taste in $X$. 
The following axiom postulates completeness when comparing alternatives without ambiguity. 

\begin{axiom}[Unambiguous Completeness]
    For all $x,y \in X$, $x\succsim y$ or $y\succsim x$. 
\end{axiom}

The next axiom postulates the transitive property only when there is an ``obvious" relation in either pair. 
Suppose $f\succsim g$. 
If another act $h$ does not yield a better outcome than that of $g$ for every state, then there is no reason to object to the evaluation that $f$ is weakly better than $h$. 
The following transitivity requires such consistency. 
Given $\succsim$, we say that an act $f$ \textit{statewisely dominates} another act $g$, denoted by  $f\ssd g$, if $f(s) \succsim g(s)$ for all $s\in S$.

\begin{axiom}[Unambiguous Transitivity]
    For all $f, g, h\in \mathcal{F}$, if either (i) $f\ssd g$ and $g \succsim h$ or (ii)  $f\succsim g$ and $g\ssd h$, then $f\succsim h$. 
\end{axiom}
 
Note that, combined with \textit{reflexivity}, this axiom implies \textit{monotonicity}---the property requiring that for all $f,g\in \mathcal{F}$, $f\ssd g$ implies $f\succsim g$.
Indeed, for $f,g\in \mathcal{F}$ such that $f\ssd g$, since $f\sim f$ holds from \textit{reflexivity}, \textit{unambiguous transitivity} implies that $f\succsim g$. 

We then introduce the axioms of continuity and independence. We omit detailed explanations because they are standard axioms. 

\begin{axiom}[Continuity]
   For all $f,g, h\in \mathcal{F}$ with $f\succ g\succ h$, there exist $\alpha, \beta \in (0,1)$ such that $\alpha f + (1 - \alpha ) h \succ g \succ \beta  f + (1 - \beta) h$. 
\end{axiom}


\begin{axiom}[Independence]
\label{axm:ind}
    For all $f,g, h\in \mathcal{F}$ and $\alpha \in (0,1)$, 
    \begin{equation*}
        f \succsim g \iff \alpha f + (1  - \alpha) h \succsim \alpha g + (1  - \alpha) h. 
    \end{equation*}
\end{axiom}

\citet{lehrer2011justifiable} provided a representation theorem for preferences satisfying these axioms. 
Let $\mathcal{K} (\Delta (S))$ be the collection of nonempty compact convex subsets of $\Delta (S)$, endowed with the Hausdorff topology.
We call a nonempty compact subset $\mathbb{P}$ of $\mathcal{K} (\Delta (S))$ a \textit{\textbf{belief collection}}. 
The following is a characterization result from \citet{nakamura2025cautiousdualselfexpectedutility}, which elaborates on Theorem 2 of \citet{lehrer2011justifiable}. 

\begin{theorem}[Theorem 4 of \citet{nakamura2025cautiousdualselfexpectedutility}; Theorem 2 of \citet{lehrer2011justifiable}]
    \label{thm:benchmark}
    A binary relation $\succsim$ over $\F$ satisfies Axioms \ref{axm:nontri}--\ref{axm:ind} if and only if there exist a nonconstant affine function $u: X\to \bR$ and a belief collection $\mathbb{P}$ such that for all $f,g\in\F$, 
    \begin{equation}
    \label{eq:def_gen}
        f\succsim g \iff \max_{\cP\in \bP} \min_{p \in \cP} \{ \mathbb{E}_p [u(f)] - \mathbb{E}_p [u(g)] \}  \geq 0. 
    \end{equation}
\end{theorem}

We call preferences represented as \eqref{eq:def_gen} \textbf{\textit{generalized Bewley preferences}}.
These preferences can be interpreted as decision criteria of a DM with multiple selves in mind. 
Each self has an imprecise prediction as a set $\cP$ of beliefs and considers $f$ to be at least as good as $g$ if the expected utility of $f$ is no less than that of $g$ in all their beliefs. 
This process corresponds to the min part in  \eqref{eq:def_gen}.
Based on selves' opinions, the DM concludes $f$ to be weakly preferable to $g$ if at least one of them evaluates them as such, which is represented by the max part in \eqref{eq:def_gen}.
In summary, each self in the DM's mind first applies the \textbf{for-all rule}, and then the DM makes a decision according to the \textbf{for-some rule}. 

Then, we present two remarks on generalized Bewley preferences.
First, it should be noted that the preferences represented as \eqref{eq:def_gen} can be rewritten using the min-of-max operator. That is, there exists a nonconstant affine function $u: X\to \bR$ and a belief collection $\mathbb{Q}$ such that for all $f,g\in\F$, 
\begin{equation*}
    f\succsim g \iff \min_{\cQ\in \bQ} \max_{q \in \cQ} \{ \mathbb{E}_q [u(f)] - \mathbb{E}_q [u(g)] \}  \geq 0. 
\end{equation*}
This representation can be interpreted as follows: First, each self with an imprecise prediction $\cQ$ compares two acts $f$ and $g$ and concludes that $f$ is weakly better than $g$ if, for some plausible belief, the expected utility of $f$ is no less than that of $g$. Based on the selves' opinions, the DM in turn makes a conclusion using the unanimity rule. 
Therefore, in this representation, each self first applies the \textbf{for-some rule}, and then the DM compares acts according to the \textbf{for-all rule}. 

Note that the belief collections $\bP$ and $\bQ$ in the max-of-min and min-of-max operators are different in general. 
Thus, when fixing a belief collection $\bP$, the ranking over acts calculated by the max-of-min operator can differ from that calculated by the min-of-max operator.



Second, generalized Bewley preferences include two well-known special cases: Bewley preferences (\citet{bewley2002knightian}) and justifiable preferences (\citet{lehrer2011justifiable}).
We say that $\succsim$ is a \textit{Bewley preference} if there exist  a nonconstant affine function $u: X\to \bR$ and a set $\cP \in \mathcal{K} (\Delta (S))$ such that for all $f,g\in \mathcal{F}$,
\begin{equation*}
    f\succsim g \iff \min_{p\in \cP} \{ \mathbb{E}_p [u(f)] - \mathbb{E}_p [u(g)] \} \geq 0.
\end{equation*}
Indeed, if $\bP=\{\cP\}$, the generalized Bewley preference $(u,\bP)$ can be viewed as the Bewley preference $(u,\cP)$.
While Bewley preferences satisfy \textit{transitivity}, they violate \textit{completeness} in general.
The other special cases are justifiable preferences studied in \citet{lehrer2011justifiable}. 
Formally, we say that $\succsim$ is a \textit{justifiable preference} if there exist a nonconstant affine function $u: X\to \bR$ and a set $\cP \in \mathcal{K} (\Delta (S))$ such that for all $f,g\in \mathcal{F}$,
\begin{equation*}
    f\succsim g \iff \max_{p\in \cP} \{ \mathbb{E}_p [u(f)] - \mathbb{E}_p [u(g)] \} \geq 0.
\end{equation*}
If $\bP = \{ \{p\}\mid p\in\cP\}$ for some $\cP\in\mathcal{K} (\Delta (S))$, the generalized Bewley preference $(u,\bP)$ coincides with the justifiable preference $(u,\cP)$.
In contrast to Bewley preferences, justifiable preferences satisfy \textit{completeness} but violate \textit{transitivity}.

\section{Rationality and Cognitive Process}
\label{sec:main}

As mentioned above, generalized Bewley preferences may violate both \textit{completeness} and \textit{transitivity}, whereas Bewley and justifiable preferences always satisfy at least one of them. This suggests that there are classes of preferences between generalized Bewley preferences and these two extreme cases that partially satisfy rationality axioms. Yet the analysis of such classes has remained largely unexplored. In this section, we examine axioms related to rationality and derive new classes of representations for these partially rational preferences. 
Our characterization theorems show that the degree of rationality is closely related to how the DM processes ambiguity.

\subsection{Completeness}

First, we examine the implication of \textit{completeness} in generalized Bewley preferences.
This axiom requires that the DM be able to compare any pair of acts and is a necessary condition for preferences to be represented by utility functions. 
In the following theorem, we characterize the implication of \textit{completeness} under generalized Bewley preferences. 

\begin{theorem}
\label{thm:uni}
     A binary relation $\succsim$ is a generalized Bewley preference and satisfies \textit{completeness} if and only if there exist a nonconstant affine function $u: X\to \bR$ and a belief collection $\mathbb{P}$ such that for all $f,g\in\F$, 
    \begin{equation}
    \label{eq:def_uni}
        f \succsim g \iff  \max\qty{ \max_{\cP\in \bP} \min_{p \in \cP} \{ \mathbb{E}_p [u(f)] - \mathbb{E}_p [u(g)] \}, ~ \min_{\cP\in \bP} \max_{p \in \cP} \{ \mathbb{E}_p [u(f)] - \mathbb{E}_p [u(g)] \} } \geq 0. 
    \end{equation}
\end{theorem}

We call preferences represented as \eqref{eq:def_uni} \textit{\textbf{disjunctive generalized Bewley preferences}}. 
In the max-of-min part, the DM compares two acts in the same way as in Theorem \ref{thm:benchmark}. 
The DM first applies the for-all rule to each prior set $\cP$ in $\bP$, as in the Bewley model, and then aggregates evaluations among the prior sets by the for-some rule. 
The min-of-max part represents the ``dual'' comparison, in which the DM first applies the for-some rule in each prior set, as in the justifiable model, and then uses the for-all rule. 
In \eqref{eq:def_uni}, the DM evaluates $f$ to be weakly better than $g$ if at least one of the max-of-min or min-of-max comparisons reaches such a conclusion. 

Note that for any $(u, \mathbb{P})$, the disjunctive generalized Bewley preference $(u, \mathbb{P})$ can compare more pairs of acts than the generalized Bewley preference $(u, \mathbb{P})$ since the former can reach a weak preference if the min-of-max evaluation yields such a conclusion.
Theorem \ref{thm:uni} shows that within the class of generalized Bewley preferences, considering the
disjunction of the two dual processes can ensure the decisiveness that follows from completeness.

Justifiable preferences are special cases of 
disjunctive generalized Bewley preferences: If $\bP = \{ \cP\}$, then \eqref{eq:def_uni} can be rewritten as 
\begin{align*}
    f\succsim g
    &\iff \max \qty{  \min_{p \in \cP} \{ \mathbb{E}_p [u(f)] - \mathbb{E}_p [u(g)] \}, ~ \max_{p \in \cP} \{ \mathbb{E}_p [u(f)] - \mathbb{E}_p [u(g)] \} } \geq 0 \\
    &\iff  \max_{p \in \cP} \{ \mathbb{E}_p [u(f)] - \mathbb{E}_p [u(g)] \} \geq 0. 
\end{align*}
Furthermore, if there exists $\cP \in \mathcal{K} (\Delta (S))$ such that $\bP = \{ \{p\} \mid p\in \cP \}$, then \eqref{eq:def_uni} is equivalent to
\begin{align*}
    f\succsim g
    &\iff \max \qty{  \max_{ \{p\} \in \bP } \{ \mathbb{E}_p [u(f)] - \mathbb{E}_p [u(g)] \}, ~ \min_{ \{p\} \in \bP} \{ \mathbb{E}_p [u(f)] - \mathbb{E}_p [u(g)] \} } \geq 0 \\
    &\iff  \max_{ \{p\} \in \bP} \{ \mathbb{E}_p [u(f)] - \mathbb{E}_p [u(g)] \} \geq 0\\ 
    & \iff  \max_{p \in \cP} \{ \mathbb{E}_p [u(f)] - \mathbb{E}_p [u(g)] \} \geq 0, 
\end{align*}
which means that $\succsim$ is a justifiable preference. 
On the other hand, a Bewley preference is in this class if and only if it is an SEU preference. 

To understand the role of \textit{completeness} in more detail, 
we provide the graphical intuition underlying Theorem \ref{thm:uni}.\footnote{
While the formal proof establishes the theorem with simple algebra, the important implication of \textit{completeness} can be understood more directly through visualization. The argument here is aimed at illuminating the key geometrical insights that are less apparent in the algebraic approach.
}
By a standard argument, we can show that there is a nonconstant affine function $u: X\to \bR$ such that for all $x,y \in X$, $x\succsim y$ if and only if $u(x) \geq u(y)$. 
Then, we can identify each act $f$ with the element $u(f)$ of the Euclidean space.
Assume without loss of generality that for some $x_0 \in X$, $u(x_0) = 0 \in \text{int} ~u(X)$. 
By the same argument in the proof of Theorem 2 of \citet{lehrer2011justifiable}, we can see that \textit{independence} implies that there exists a nonempty closed cone $C \subset \mathbb{R}^S$ such that for all $f,g\in\F$, $f\succsim g$ if and only if $u(f) - u(g) \in C$; that is, only differences in utility vectors determine the rankings. 
\citeauthor{lehrer2011justifiable} obtained their characterization result by showing that $C$ can be represented by a union of convex cones, which corresponds to the desired max-of-min form. 

\begin{figure}
    \centering
    
    
    \includegraphics[width=1.0\linewidth]{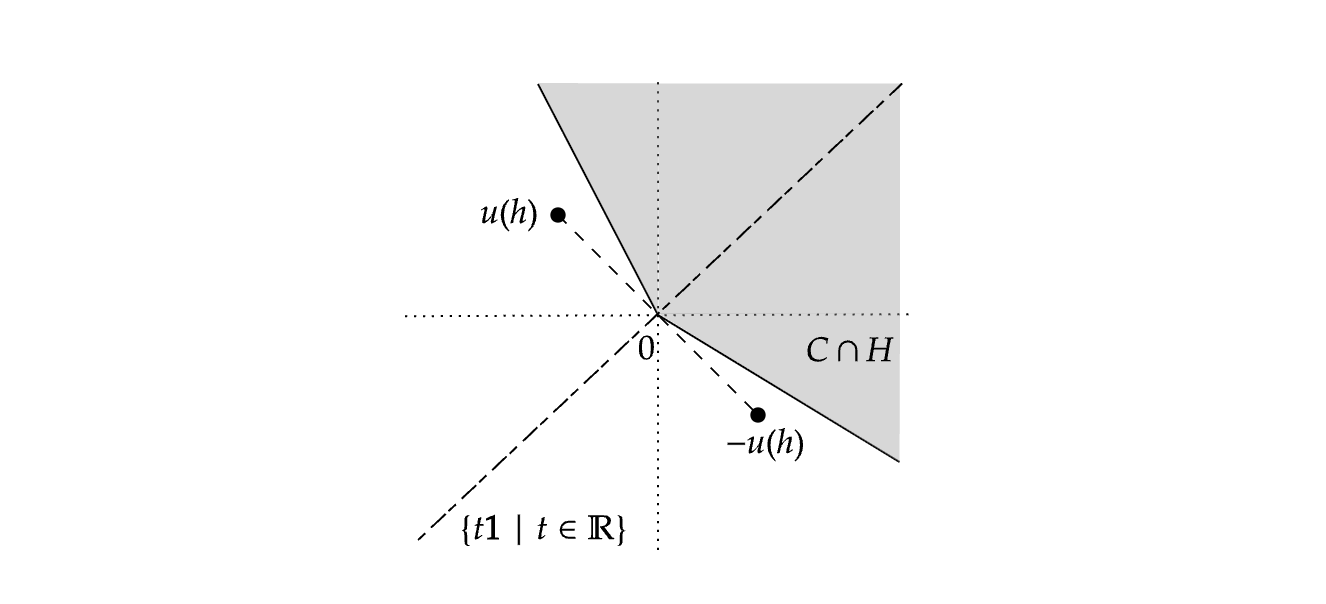}
    \caption{$u(h)$ on $H$}
    \label{fig_supp1_thmuni}
\end{figure}

We say that a nonempty cone in $\mathbb{R}^S$ is \textbf{\textit{diagonal-slice convex}} if, for each two-dimensional plane $H \subset \mathbb{R}^S$ that includes the diagonal line $\{ t \mathbf{1} \mid t\in \mathbb{R} \}$, the intersection of this cone and $H$ is a convex set. 
The geometric implication of \textit{completeness} is that the complement of the cone $C$ is diagonal-slice convex. 
To see this, suppose to the contrary that the complement of the cone $C$ is not diagonal-slice convex.
Then, we can take $h\in \F$ such that $u(h) , -u(h)\notin C$ (see Figure \ref{fig_supp1_thmuni}).
This implies that 
 $u(h) -u(x_0) , u(x_0)-u(h)\notin C$. 
By definition, \textit{completeness} implies that $u(h) -u(x_0) \in C $ or $u(x_0) - u(h) \in C$ must hold, 
 which is a contradiction.

\begin{figure}
    \centering 
    
    \begin{subfigure}[b]{0.48\textwidth} 
    \centering

    
    \begin{tikzpicture}[x=0.75pt,y=0.75pt,yscale=-1,xscale=1]
    
    \draw  [dash pattern={on 0.84pt off 2.51pt}]  (202.5,158.12) -- (455.5,157.12) ;
    \draw  [dash pattern={on 0.84pt off 2.51pt}]  (329,275) -- (329,38) ;
    \draw    (285,43) -- (329,157.5) ;
    \draw    (329,157.5) -- (455,200) ;
    \draw [color={rgb, 255:red, 208; green, 2; blue, 27 }  ,draw opacity=1 ]   (203,115.12) -- (329,157.62) ;
    \draw [color={rgb, 255:red, 208; green, 2; blue, 27 }  ,draw opacity=1 ]   (329,156.5) -- (373,271) ;
    \draw  [dash pattern={on 3.75pt off 3pt on 7.5pt off 1.5pt}]  (210,270) -- (453,42) ;
    
    \draw (321,166.4) node [anchor=north west][inner sep=0.75pt]    {$0$};
    \draw (401,162.4) node [anchor=north west][inner sep=0.75pt]    {$K_{\mathbb{P}} \cap H$};
    \draw (379,244.4) node [anchor=north west][inner sep=0.75pt]    {$K_{\mathbb{P}}^{*} \cap H$};
    \draw (228,256.4) node [anchor=north west][inner sep=0.75pt]    {$\{t\mathbf{1} \mid t\in \mathbb{R}\}$};

    \end{tikzpicture}

        \caption{$K_\mathbb{P} \cap H$ is convex}
        \label{fig_supp2_cc_a}
    \end{subfigure}
    \hfill 
    \begin{subfigure}[b]{0.48\textwidth} 
        \centering
        \begin{tikzpicture}[x=0.75pt,y=0.75pt,yscale=-1,xscale=1]

\draw  [dash pattern={on 0.84pt off 2.51pt}]  (202.5,158.12) -- (455.5,157.12) ;
\draw  [dash pattern={on 0.84pt off 2.51pt}]  (329,275) -- (329,38) ;
\draw [color={rgb, 255:red, 208; green, 2; blue, 27 }  ,draw opacity=1 ]   (285,43) -- (329,157.5) ;
\draw [color={rgb, 255:red, 208; green, 2; blue, 27 }  ,draw opacity=1 ]   (329,157.5) -- (455,200) ;
\draw [color={rgb, 255:red, 0; green, 0; blue, 0 }  ,draw opacity=1 ]   (203,115.12) -- (329,157.62) ;
\draw [color={rgb, 255:red, 0; green, 0; blue, 0 }  ,draw opacity=1 ]   (329,156.5) -- (373,271) ;
\draw  [dash pattern={on 3.75pt off 3pt on 7.5pt off 1.5pt}]  (207.5,270.5) -- (450.5,42.5) ;

\draw (320,166.4) node [anchor=north west][inner sep=0.75pt]    {$0$};
\draw (401,162.4) node [anchor=north west][inner sep=0.75pt]    {$K_{\mathbb{P}}^{*} \cap H$};
\draw (379,244.4) node [anchor=north west][inner sep=0.75pt]    {$K_{\mathbb{P}} \cap H$};
\draw (230,255.4) node [anchor=north west][inner sep=0.75pt]    {$\{t\mathbf{1} \mid t\in \mathbb{R}\}$};

\end{tikzpicture}
        \caption{$K^*_\mathbb{P} \cap H$ is convex}
        \label{fig_supp2_cc_b}
    \end{subfigure}
    
    \caption{Diagonal-slice convexity and the union of $K_\mathbb{P}$ and $K^*_\mathbb{P}$}
    \label{fig_supp2_cc}
\end{figure}
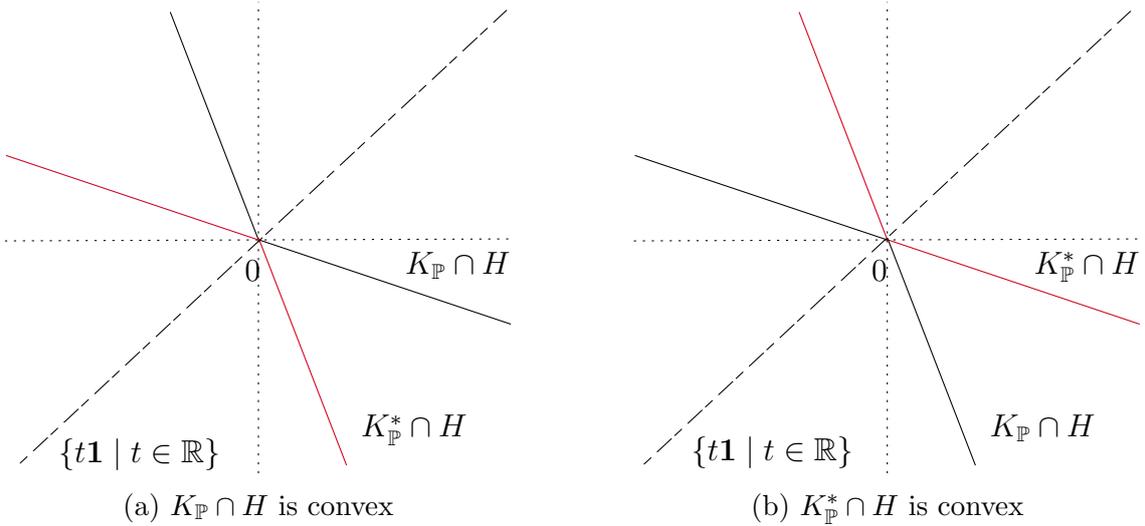

Then, to illustrate the intuition behind Theorem \ref{thm:uni}, we discuss how the diagonal-slice convexity of the complement of $C$ and the disjunctive generalized Bewley preferences are connected.
Fix a belief collection $\mathbb{P}$. Consider the cones 
\begin{equation*}
    K_\mathbb{P} = \qty{ \varphi\in \mathbb{R}^S \mid \max_{\cP \in \bP} \min_{p\in \cP} \mathbb{E}_p [\varphi] \geq 0 } \quad \text{and} \quad K^*_\mathbb{P} = \qty{ \varphi\in \mathbb{R}^S \mid \min_{\cP \in \bP} \max_{p\in \cP} \mathbb{E}_p [\varphi] \geq 0 }. 
\end{equation*}
Note that $K_\mathbb{P}\cup K^*_\mathbb{P}$ corresponds to the upper contour set of the disjunctive generalized Bewley preferences associated with $\mathbb{P}$. 
We now show that the complement of $K_\mathbb{P}\cup K^*_\mathbb{P}$ is diagonal-slice convex.
Take an arbitrary two-dimensional plane $H$ that includes the diagonal line. 
Since $K_\bP$ is a cone, either $K_\bP$ or its complement is convex on $H$.
First, consider the case where $K_\bP$ on $H$ is convex (Figure \hyperref[fig_supp2_cc_a]{2(a)}). 
Note that the boundary of $K_\bP^*$ can be obtained by rotating that of $K_\bP$ by 180 degrees about the origin. 
In this case,  $K_\mathbb{P}\cup K^*_\mathbb{P}$ on $H$ is equal to $K_\bP^*$ on $H$, so the complement of  $K_\mathbb{P}\cup K^*_\mathbb{P}$ on $H$ is convex. 
Similarly, if the complement of $K_\bP$ is convex on $H$, then $K_\mathbb{P}\cup K^*_\mathbb{P}$ coincides with $K_\bP$ on $H$ (Figure \hyperref[fig_supp2_cc_b]{2(b)}). Again, the complement of  $K_\mathbb{P}\cup K^*_\mathbb{P}$ on $H$ is convex, as desired. 
Therefore, both imposing \textit{completeness} and taking the union of the max-of-min and min-of-max procedures are closely linked to the diagonal-slice convexity of the complement of upper contour sets. 

Finally, we note the difference between the disjunctive generalized Bewley preferences and the justifiable preferences at the axiomatic level. 
\citet{lehrer2011justifiable} characterized the justifiable preferences by additionally imposing the following axiom on disjunctive generalized Bewley preferences: 

\begin{axiom}[Favorable Mixing]
    For all $f,g,h \in \F$ and $\alpha, \lambda  \in (0,1)$ with $\lambda \leq \alpha$, if $g\succ f$ and $\alpha f + (1-\alpha ) h \succsim g$, then  $\lambda f + (1-\lambda ) h \succsim g$. 
\end{axiom}

This axiom postulates that if $f$ is strictly worse than $g$ and if we can obtain an act weakly better than $g$ by mixing $f$ and some act $h$ with proportion $\alpha: 1-\alpha$, then every mixed act assigning a higher proportion to $h$ should also be weakly better than $g$. 
This requirement is very reasonable, but the implication of this axiom \textit{per se} was not explored by \citet{lehrer2011justifiable}. 

Theorem \ref{thm:uni} considers the axioms used in the characterization of justifiable preferences except for \textit{favorable mixing}. 
Thus, by comparing these two characterizations, we can see how \textit{favorable mixing} works in \citeauthor{lehrer2011justifiable}'s characterization. 
In their proof, \textit{favorable mixing} ensures that the complement of $C$ in the above graphical explanation is convex, which is a much stronger condition than diagonal-slice convexity implied by \textit{completeness}.

\subsection{Weak Transitivity}


While \textit{transitivity} is often viewed as a cornerstone of rational choice, imposing it on generalized Bewley preferences reduces them to the traditional Bewley model.  
This raises a natural question: Is there a weaker condition that still captures meaningful consistency without being as restrictive as full transitivity? 
To address this, we introduce a mild condition that links the evaluation of ambiguous acts to their certain counterparts.

\begin{axiom}[Constant-Bound Transitivity]
    For all $f\in \F$ and $x,y\in X$, if $x\succsim f$ and $f\succsim y$, then $x\succsim y$. 
\end{axiom}
 
This postulates that constant acts can be used to approximate the value of any act.\footnote{
Axiom 5 of \citet{echenique2022twofold} requires the following transitive property: For all $f,g\in \F$ and $x\in X$, if $f$ and $g$ are incomparable with $x$, then $f$ and $g$ should also be incomparable. 
Similar to \textit{constant-bound transitivity}, this axiom is motivated by the idea that constant acts serve as baseline evaluations.
} 
Since the DM knows the value of constant acts, $x \succsim f$ and $f \succsim y$ can be interpreted as offering unambiguous upper and lower caps of the value of $f$, respectively. 
\textit{Constant-bound transitivity} postulates that the reversal of these caps (i.e., $y\succ x$) does not occur, which means the value of $f$ can be consistently approximated using constant acts. 

If this axiom is violated, then $x\succsim f  \succsim y$ but $y\succ x$ hold under \textit{unambiguous completeness}. 
Suppose that $x$ and $y$ are alternatives that yield \$100 and \$200, respectively,  with certainty (note that this is consistent with $y\succ x$). 
Then $x\succsim f  \succsim y$ means that $f$ is weakly preferable to \$200 but weakly worse than \$100. 
\textit{Constant-bound transitivity} excludes such ill-behaved preference patterns. 

Furthermore, \textit{constant-bound transitivity} implies that if $x\sim f$ and $y\sim f$, then $x\sim y$. 
Thus, this axiom implies the natural property that if there are two certainty equivalents, then they are equally desirable. 


The following result shows that \textit{constant-bound transitivity} has the dual implication of \textit{completeness}. 

\begin{theorem}
\label{thm:int}
    A  binary relation $\succsim$ is a generalized Bewley preference and satisfies \textit{constant-bound transitivity} if and only if  there exist a nonconstant affine function $u: X\to \bR$ and a belief collection $\mathbb{P}$ such that for all $f,g\in\F$, 
     \begin{equation}
    \label{eq:def_inter}
        f \succsim g \iff  \min\qty{ \max_{\cP\in \bP} \min_{p \in \cP} \{ \mathbb{E}_p [u(f)] - \mathbb{E}_p [u(g)] \}, ~ \min_{\cP\in \bP} \max_{p \in \cP} \{ \mathbb{E}_p [u(f)] - \mathbb{E}_p [u(g)] \} } \geq 0. 
    \end{equation}
\end{theorem}

We call preferences represented as \eqref{eq:def_inter} \textit{\textbf{conjunctive generalized Bewley preferences}}. 
In these preferences, the DM evaluates $f$ to be weakly better than $g$ if both of the max-of-min and min-of-max comparisons conclude so. 
Compared with the disjunctive generalized Bewley preferences, the DM uses the two comparisons in the opposite way. 

As discussed in Section \ref{sec:intro}, \textit{(constant-bound) transitivity} has the effect opposite to \textit{completeness}; that is, the DM cannot reach a weak relation easily because \textit{(constant-bound) transitivity} prohibits inconsistent choices. 
For any $(u, \mathbb{P})$, the conjunctive generalized Bewley preference $(u, \mathbb{P})$ 
is less likely to conclude that one act is weakly preferred to another than the generalized Bewley preference $(u, \mathbb{P})$. 
Therefore, the conjunctive generalized Bewley preference $(u, \mathbb{P})$ is less likely to violate \textit{transitivity}. 
Theorem \ref{thm:int} formally establishes the connection between \textit{constant-bound transitivity} and cognitive processes. 

Note that Bewley preferences are special cases of conjunctive generalized Bewley preferences: If $\bP = \{ \cP\}$, then \eqref{eq:def_inter} can be rewritten as 
\begin{align*}
    f\succsim g
    &\iff \min\qty{  \min_{p \in \cP} \{ \mathbb{E}_p [u(f)] - \mathbb{E}_p [u(g)] \}, ~ \max_{p \in \cP} \{ \mathbb{E}_p [u(f)] - \mathbb{E}_p [u(g)] \} } \geq 0 \\
    &\iff  \min_{p \in \cP} \{ \mathbb{E}_p [u(f)] - \mathbb{E}_p [u(g)] \} \geq 0. 
\end{align*}
Furthermore, if there exists $\cP \in \mathcal{K} (\Delta (S))$ such that $\bP = \{ \{p\} \mid p\in \cP \}$, then \eqref{eq:def_inter} is equivalent to
\begin{align*}
    f\succsim g
    &\iff \min\qty{  \max_{ \{p\} \in \bP } \{ \mathbb{E}_p [u(f)] - \mathbb{E}_p [u(g)] \}, ~ \min_{ \{p\} \in \bP} \{ \mathbb{E}_p [u(f)] - \mathbb{E}_p [u(g)] \} } \geq 0 \\
    &\iff  \min_{ \{p\} \in \bP} \{ \mathbb{E}_p [u(f)] - \mathbb{E}_p [u(g)] \} \geq 0\\ 
    & \iff  \min_{p \in \cP} \{ \mathbb{E}_p [u(f)] - \mathbb{E}_p [u(g)] \} \geq 0, 
\end{align*}
which means that $\succsim$ is a Bewley preference. 
On the other hand, a justifiable preference is in this class if and only if it is an SEU preference. 

We then illustrate the geometric connection between \textit{constant-bound transitivity} and conjunctive generalized Bewley preferences, thereby clarifying the intuition behind Theorem \ref{thm:int}. 
First, note that, as explained in the previous part, there exist a nonconstant affine function $u: X\to \mathbb{R}$ and a nonempty closed cone $C \subset \mathbb{R}^S$ such that for all $f,g\in\F$, $f\succsim g$ if and only if $u(f) - u(g) \in C$. 
We take $x_0 \in X$ such that $u(x_0) = 0 \in \text{int} ~ u(X)$ without loss of generality. 
By \textit{monotonicity}, $C$ includes $\mathbb{R}^S_{+}$ and does not meet $\mathbb{R}^S_{--}$.\footnote{Let $\mathbb{R}_{--}$ denote the set of negative numbers. }

\begin{figure}
    \centering
    \includegraphics[width=1.0\linewidth]{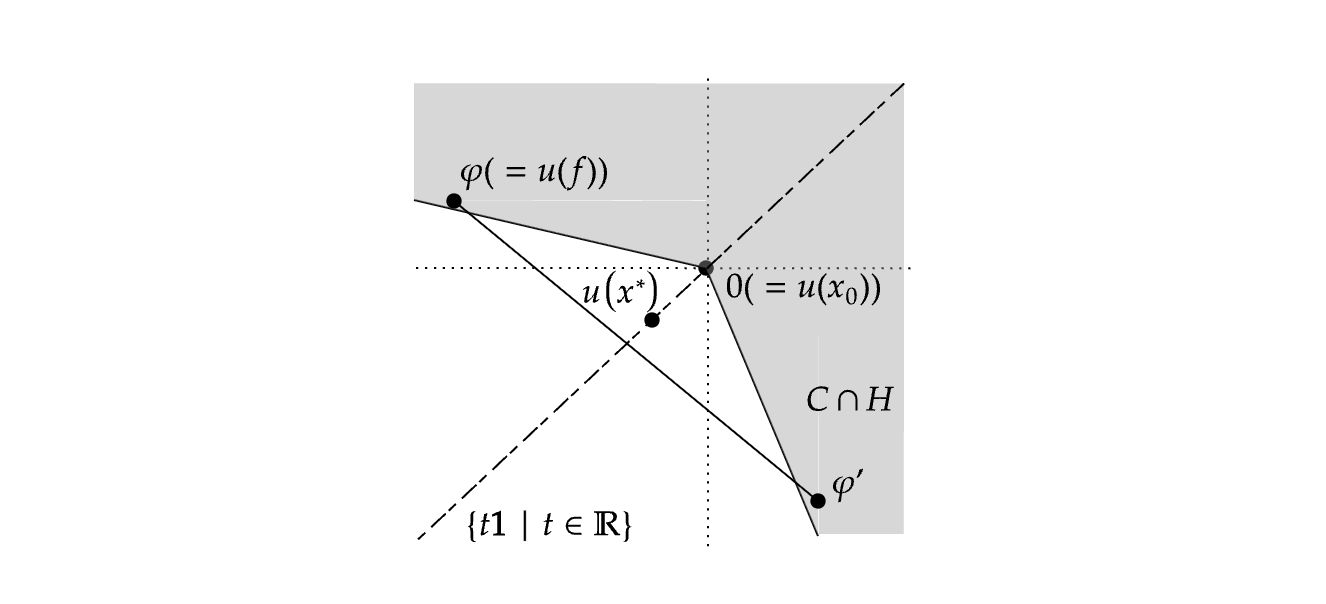}
    \caption{$u(f)$ and $u(x^*)$}
    \label{fig:CBtrans}
\end{figure}

The key implication of \textit{constant-bound transitivity} is that $C$ becomes diagonal-slice convex. 
To see this, suppose instead that for some two-dimensional plane $H \subset \mathbb{R}^S$ that includes the diagonal line $\{ t \mathbf{1} \mid t\in \mathbb{R} \}$, $C\cap H$ is not a convex set. 
Then, since $C\cap H$ is also a cone, 
we can take $\varphi, \varphi' \in C$ such that for some $t^* < 0$, $ \varphi + \varphi'  = t^* \mathbf{1} \notin C$ (Figure \ref{fig:CBtrans}). 
For $\varepsilon > 0$ with $t^* + \varepsilon < 0 $ (i.e., $ (t^* + \varepsilon )\mathbf{1} \notin C$), we have $ (t^* + \varepsilon )\mathbf{1} -\varphi = \varphi' + \varepsilon \mathbf{1} \in C $. 
Consider $f \in \F$ and $x^* \in X$ such that $u(f) = \varphi$ and $u(x^*) = t^* + \varepsilon$. 
Since $u(f) - u(x_0) = \varphi \in C$ and $u(x^*) - u(f) = (t^* + \varepsilon )\mathbf{1} -\varphi \in C$, we have $f\succsim x_0$ and $x^* \succsim f$. 
However, since $u(x^*) = t^* + \varepsilon < 0 = u(x_0)$, we have $x^* \not\succsim x_0$, which contradicts \textit{constant-bound transitivity}.

Finally, we explain the relationship between the diagonal-slice convexity of $C$ and the intersection of the max-of-min form and min-of-max form. Fix a belief collection $\mathbb{P}$, and again consider the cones 
\begin{equation*}
    K_\mathbb{P} = \qty{ \varphi\in \mathbb{R}^S \mid \max_{\cP \in \bP} \min_{p\in \cP} \mathbb{E}_p [\varphi] \geq 0 } \quad \text{and} \quad K^*_\mathbb{P} = \qty{ \varphi\in \mathbb{R}^S \mid \min_{\cP \in \bP} \max_{p\in \cP} \mathbb{E}_p [\varphi] \geq 0 }. 
\end{equation*}
We can show that $K_\mathbb{P}\cap K^*_\mathbb{P}$ is diagonal-slice convex.
Take an arbitrary two-dimensional plane $H$ that includes the diagonal line. 
Since $K_\bP$ is a cone, either $K_\bP$ or its complement is convex on $H$.
First, consider the case where $K_\bP$ is convex (Figure \hyperref[fig_supp2_cc_a]{2(a)}). Note that the boundary of $K_\bP^*$ can be obtained by rotating that of $K_\bP$ by 180 degrees about the origin. 
In this case,  $K_\mathbb{P}\cap K^*_\mathbb{P}$ on $H$ is equal to $K_\bP$ on $H$, so $K_\mathbb{P}\cap K^*_\mathbb{P}$ on $H$ is convex. 
Similarly, for the case where the complement of $K_\bP$ is convex on $H$,  $K_\mathbb{P}\cap K^*_\mathbb{P}$ coincides with $K_\bP^*$ on $H$ (Figure \hyperref[fig_supp2_cc_a]{2(b)}). 
Again, $K_\mathbb{P}\cap K^*_\mathbb{P}$ on $H$ is convex, as desired. 
Therefore, both imposing \textit{constant-bound transitivity} and taking the intersection of the max-of-min and min-of-max procedures are closely linked to the diagonal-slice convexity of the upper contour set. 
This relationship is the intuition behind Theorem \ref{thm:int}.

\subsection{Joint Implication}

This subsection examines the joint implication of \textit{completeness} and \textit{constant-bound transitivity}. 
The generalized Bewley preferences that satisfy these two axioms can be characterized as follows.  

\begin{theorem}
\label{thm_half}
     A  binary relation $\succsim$ is a generalized Bewley preference and satisfies \textit{completeness} and \textit{constant-bound transitivity} if and only if  there exist a nonconstant affine function $u: X\to \bR$ and a belief collection $\mathbb{P}$ such that for all $f,g\in\F$, 
     \begin{equation}
     \label{eq:def_half}
        f \succsim g \iff  {1\over 2}\max_{\cP\in \bP} \min_{p \in \cP} \{ \mathbb{E}_p [u(f)] - \mathbb{E}_p [u(g)] \} + {1\over 2} \min_{\cP\in \bP} \max_{p \in \cP} \{ \mathbb{E}_p [u(f)] - \mathbb{E}_p [u(g)] \geq 0. 
    \end{equation}
\end{theorem}

We call preferences represented as \eqref{eq:def_half} \textit{\textbf{half-mixture generalized Bewley preferences}}. 
As in the two previous models, half-mixture generalized Bewley preferences consider two ways of evaluation---the max-of-min and min-of-max comparisons. 
This class uses both comparisons equally for each pair of acts. 
The first term can be considered the benefit of $f$ relative to $g$ under the max-of-min comparison, and the second term represents the relative benefit of $f$ with respect to $g$  under the min-of-max comparison. 
A DM with the preference \eqref{eq:def_half} evaluates $f$ to be weakly better than $g$ if the average  benefit of $f$ relative to $g$ is nonnegative.

As discussed, \textit{completeness} and \textit{constant-bound transitivity} have the opposite effects on decisiveness. 
Theorem \ref{thm_half} shows that the half-mixture operation can balance the opposite effects while ensuring these two axioms.


To understand the geometric intuition in Theorem \ref{thm_half}, recall that for a generalized Bewley preference $\succsim$, there are a nonconstant affine function $u: X\to \mathbb{R}$ and a nonempty closed cone $C$ such that $f\succsim g$ if and only if $u(f) - u(g) \in C$. 
As illustrated in the previous subsections, \textit{completeness} implies that the complement of $C$ is diagonal-slice convex, 
and \textit{constant-bound transitivity} implies that $C$ is so. 
Thus, for each two-dimensional plane $H$ that includes the diagonal line, $C\cap H$ is a half space under these two axioms.

\begin{figure}
    \centering
    \begin{tikzpicture}[x=0.75pt,y=0.75pt,yscale=-1,xscale=1]
    
    \draw  [dash pattern={on 0.84pt off 2.51pt}]  (202.5,158.12) -- (455.5,157.12) ;
    \draw  [dash pattern={on 0.84pt off 2.51pt}]  (329,275) -- (329,38) ;
    \draw    (232.5,41) -- (329,157.5) ;
    \draw    (329,157.5) -- (453.5,189) ;
    \draw [color={rgb, 255:red, 208; green, 2; blue, 27 }  ,draw opacity=1 ]   (202.5,119) -- (329,157.62) ;
    \draw [color={rgb, 255:red, 208; green, 2; blue, 27 }  ,draw opacity=1 ]   (329,157.5) -- (422.5,274.5) ;
    \draw  [dash pattern={on 3.75pt off 3pt on 7.5pt off 1.5pt}]  (210,270) -- (453,42) ;
    \draw [color={rgb, 255:red, 74; green, 144; blue, 226 }  ,draw opacity=1 ][line width=1.5]    (202.5,72.5) -- (452.5,239.5) ;
    
    \draw (321,166.4) node [anchor=north west][inner sep=0.75pt]    {$0$};
    \draw (260,50.4) node [anchor=north west][inner sep=0.75pt]    {$K_{\mathbb{P}} \cap H$};
    \draw (409,241.4) node [anchor=north west][inner sep=0.75pt]    {$K_{\mathbb{P}}^{*} \cap H$};
    \draw (228,256.4) node [anchor=north west][inner sep=0.75pt]    {$\{t\mathbf{1} \mid t\in \mathbb{R}\}$};
    \draw (424,198.4) node [anchor=north west][inner sep=0.75pt]    {$K_{\mathbb{P}}^{\text{half}}$};
    
    \end{tikzpicture}
    \caption{$K^\text{half}_\bP$ on $H$}
    \label{fig:half}
\end{figure}
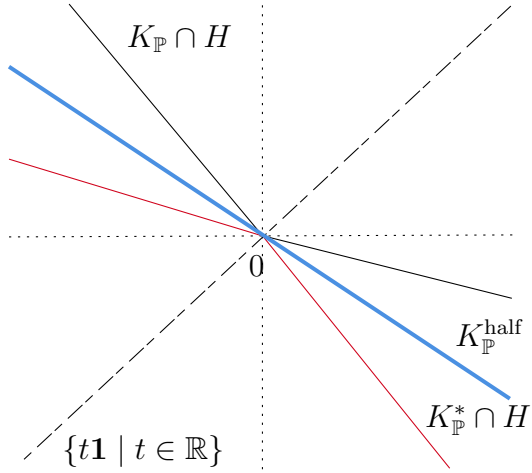

Then, we explain the relationship between the above geometric property and the half-mixture operation. 
Take an arbitrary belief collection $\bP$ and consider the two cones
\begin{equation*}
    K_\mathbb{P} = \qty{ \varphi\in \mathbb{R}^S \mid \max_{\cP \in \bP} \min_{p\in \cP} \mathbb{E}_p [\varphi] \geq 0 } 
    \quad \text{and} 
    \quad K^*_\mathbb{P} = \qty{ \varphi\in \mathbb{R}^S \mid \min_{\cP \in \bP} \max_{p\in \cP} \mathbb{E}_p [\varphi] \geq 0 }. 
\end{equation*}
Note that the boundary of $K_\bP^*$ can be obtained by rotating that of $K_\bP$ by 180 degrees about the origin. 
Let
\begin{equation*}
    K^\text{half}_\mathbb{P} = \qty{ \varphi\in \mathbb{R}^S \mid {1\over 2} \max_{\cP \in \bP} \min_{p\in \cP} \mathbb{E}_p [\varphi] + {1\over 2}\min_{\cP \in \bP} \max_{p\in \cP} \mathbb{E}_p [\varphi]  \geq 0 }.   
\end{equation*}
For each two-dimensional plane $H$ that includes the diagonal line, the boundary of $K^\text{half}_\mathbb{P}$ can be drawn as the blue line in Figure \ref{fig:half}; that is, it is the angle bisector of the boundaries of $K_\bP$ and $K^*_\bP$ on $H$. 
Thus, by considering the half-mixture operation, we can construct a preference whose indifference curve on each two-dimensional plane $H$ that includes the diagonal line is a half space.

In addition to Theorem \ref{thm_half}, the following result holds under \textit{completeness} and \textit{constant-bound transitivity}: 

\begin{proposition}
\label{pro:interchange}
    Let $\succsim$ be a generalized Bewley preference associated with $(u, \bP)$.
    Then, $\succsim$ satisfies \textit{constant-bound transitivity} and \textit{completeness} if and only if for all $f, g\in \mathcal{F}$, 
    \begin{equation*}
         \max_{\cP\in \bP} \min_{p \in \cP} \{ \mathbb{E}_p [u(f)] - \mathbb{E}_p [u(g)] \} =  \min_{\cP\in \bP} \max_{p \in \cP} \{ \mathbb{E}_p [u(f)] - \mathbb{E}_p [u(g)] \}. 
    \end{equation*}
\end{proposition}

This proposition shows that \textit{completeness} and \textit{constant-bound transitivity} is a necessary and sufficient condition for the commutativity of the max and min operations. 
Therefore, under these two axioms, a generalized Bewley preference $\succsim$ associated with  $(u,\mathbb{P})$ can be written as, for all $f,g\in \F$, 
\begin{equation*}
    f\succsim g 
    \iff 
    \min_{P\in \bP} \max_{p\in \cP} \{ \mathbb{E}_p [u(f)] - \mathbb{E}_p [u(g)] \} \geq 0. 
\end{equation*}
In this case, it does not matter which rule should be applied first from the for-all and for-some rules. 

\begin{figure}
    \centering
    \begin{tikzpicture}[x=0.75pt,y=0.75pt,yscale=-1,xscale=1]
    
    \draw  [dash pattern={on 0.84pt off 2.51pt}]  (202.5,158.12) -- (455.5,157.12) ;
    \draw  [dash pattern={on 0.84pt off 2.51pt}]  (329,275) -- (329,38) ;
    \draw    (232.5,41) -- (329,157.5) ;
    \draw    (329,157.5) -- (453.5,189) ;
    \draw [color={rgb, 255:red, 208; green, 2; blue, 27 }  ,draw opacity=1 ]   (202.5,119) -- (329,157.62) ;
    \draw [color={rgb, 255:red, 208; green, 2; blue, 27 }  ,draw opacity=1 ]   (329,157.5) -- (422.5,274.5) ;
    \draw  [dash pattern={on 3.75pt off 3pt on 7.5pt off 1.5pt}]  (210,270) -- (453,42) ;
    \draw [color={rgb, 255:red, 74; green, 144; blue, 226 }  ,draw opacity=1 ][line width=1.5]    (329,157.62) -- (453.5,221) ;
    \draw [color={rgb, 255:red, 74; green, 144; blue, 226 }  ,draw opacity=1 ][line width=1.5]    (201.5,48) -- (329,157.62) ;
    
    \draw (321,166.4) node [anchor=north west][inner sep=0.75pt]    {$0$};
    \draw (260,50.4) node [anchor=north west][inner sep=0.75pt]    {$K_{\mathbb{P}} \cap H$};
    \draw (409,241.4) node [anchor=north west][inner sep=0.75pt]    {$K_{\mathbb{P}}^{*} \cap H$};
    \draw (228,256.4) node [anchor=north west][inner sep=0.75pt]    {$\{t\mathbf{1} \mid t\in \mathbb{R}\}$};
    \draw (431,191.4) node [anchor=north west][inner sep=0.75pt]    {$K_{\mathbb{P}}^{\alpha }\cap H$};

    \end{tikzpicture}
    \caption{The case where the weight is unequal}
    \label{fig:counterex}
\end{figure}
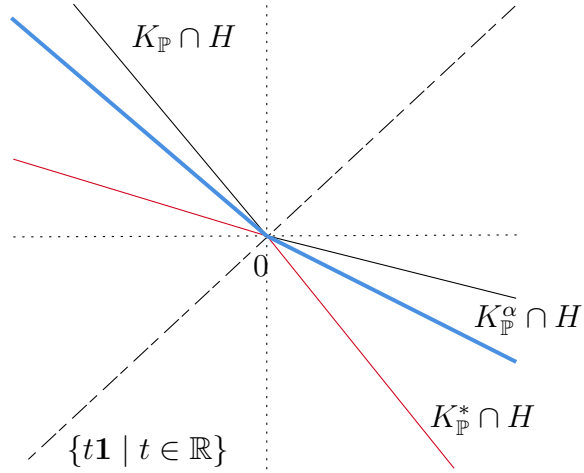

It should be noted that this proposition does not mean that we can change the weights to the max-of-min and min-of-max operations arbitrarily.
To see this, consider the preference $\succsim$ such that for all $f,g\in \F$, 
 \begin{equation*}
        f \succsim g \iff  \alpha \max_{\cP\in \bP} \min_{p \in \cP} \{ \mathbb{E}_p [u(f)] - \mathbb{E}_p [u(g)] \} + (1-\alpha) \min_{\cP\in \bP} \max_{p \in \cP} \{ \mathbb{E}_p [u(f)] - \mathbb{E}_p [u(g)] \geq 0, 
\end{equation*}
where $\alpha \in (1/2, 1)$. 
Let 
\begin{equation*}
    K^\alpha_\bP
    = 
    \qty{ \varphi\in \mathbb{R}^S \mid \alpha \max_{\cP \in \bP} \min_{p\in \cP} \mathbb{E}_p [\varphi] + (1-\alpha) \min_{\cP \in \bP} \max_{p\in \cP} \mathbb{E}_p [\varphi]  \geq 0 }.
\end{equation*}
In general, the cones  $K_\bP$ and  $K_\bP^*$ do not coincide; hence, $K^\alpha_\bP$ bends toward $K_\bP$ (for an illustration on a two-dimensional plane $H$ that includes the diagonal line, see Figure \ref{fig:counterex}). 
Thus, $\succsim$ may violate \textit{completeness} or \textit{constant-bound transitivity} in this case. 

Finally, from the above graphical argument, we can see that in the binary state case, the SEU model can be characterized by the preferences considered in Theorem \ref{thm_half}.

\begin{proposition}
\label{pro:seu}
    Suppose that $|S| = 2$. 
    Then, $\succsim$ is a generalized Bewley preference and satisfies \textit{completeness} and \textit{constant-bound transitivity} if and only if it is an SEU preference. 
\end{proposition}

According to this proposition, when $|S| = 2$, the role of \textit{transitivity} can be decomposed into  \textit{unambiguous transitivity} and \textit{constant-bound transitivity}. 

\section{Discussion}
\label{sec:alt}

\subsection{Parametric Characterizations}

So far, we have studied the novel cognitive procedures obtained from \textit{completeness} and \textit{constant-bound transitivity}.
Since these preferences are subclasses of generalized Bewley preferences, we can represent them by imposing specific restrictions on their belief collections in the generalized Bewley model.
In this subsection, we explore which properties of belief collections follow from these axioms.
By comparing the following analysis with our main results, we can see that the main theorems capture the duality of the two axioms, which is difficult to identify in parametric characterizations.

We begin with the implication of \textit{completeness}.
We say that a belief collection $\mathbb{P}$ \textit{has a cutting hyperplane} if there exists a pair $(\varphi, \lambda) \in \mathbb{R}^S \times \mathbb{R}$ such that for each $\cP\in \mathbb{P}$, there exist $p^+,p^- \in \cP$ with $ \sum_{s\in S} p^+ (s) \varphi(s)> \lambda > \sum_{s\in S} p^- (s) \varphi(s)$.
The following result shows that the implication  of \textit{completeness} can be characterized by the nonexistence of a cutting hyperplane.

\begin{proposition}
\label{prp:para_comp}
    Let $\succsim$ be a generalized Bewley preference associated with $(u, \mathbb{P})$. Then, $\succsim$ satisfies \textit{completeness} if and only if $\mathbb{P}$ does not have a cutting hyperplane. 
\end{proposition}

If $\bP$ has a cutting hyperplane, then all the belief sets in $\bP$ lie on a single line and, in that sense, the belief sets are similar.
Then, we can find a pair of acts for which unanimous judgment cannot be reached in every belief set.
Since the DM cannot compare such a pair of acts, \textit{completeness} is violated.
In this way, Proposition \ref{prp:para_comp} links the decisiveness of preferences to the nonexistence of a cutting hyperplane.
On the other hand, Theorem \ref{thm:uni} shows that decisiveness can be ensured by the union operation applied to the dual cognitive procedures.  

We next consider the implication of \textit{constant-bound transitivity} at the parameter level.
The following result shows that this axiom requires every pair in $\bP$ to intersect.

\begin{proposition}
\label{prp:para_tran}
    Let $\succsim$ be a generalized Bewley preference associated with $(u, \mathbb{P})$. Then, $\succsim$ satisfies \textit{constant-bound transitivity} if and only if $\cP \cap \cP'\neq\emptyset$ for all $\cP,\cP' \in \mathbb{P}$.
\end{proposition}

The condition that every pair in $\bP$ has an intersection ensures that the belief sets are sufficiently similar.
If there is a pair that does not satisfy this condition, one can construct an act whose evaluations differ substantially between the two.
More precisely, one belief set in $\bP$ unanimously evaluates the act as better than an outcome $x$, while another set evaluates it as worse than an outcome $y$, even though $x$ is superior to $y$.
This leads to a violation of \textit{constant-bound transitivity}.
By contrast, Theorem \ref{thm:int} shows that taking the intersection of dual processes prevents inconsistent choices and thereby guarantees \textit{constant-bound transitivity}.

Finally, we note that the conditions on belief collections established in Propositions \ref{prp:para_comp} and \ref{prp:para_tran} are not straightforward to verify. 
In particular, to check both \textit{completeness} and \textit{constant-bound transitivity}, we must consider two distinct conditions concerning intersections and cutting hyperplanes.
In contrast, our main theorems do not impose additional restrictions on belief collections, which is a practical advantage of these models.


\subsection{Negative Relation}

Our main theorems reveal a duality between two rationality conditions: \textit{Completeness} and \textit{constant-bound transitivity} each correspond to disjunctive and conjunctive generalized Bewley preferences. 
In this subsection, we show that by considering rationality with respect to the negative part of preferences, these axioms are linked to the opposite models.
These results illuminate the dual relationship between the two axioms from another perspective.

When considering the negative version of completeness, we cannot adopt the negative part $\not\succsim$ directly because requiring completeness  for this part would rule out any indifference relation.\footnote{Indeed, by 
\begin{equation}
     f\not\succsim g \iff \max_{\cP \in \bP} \min_{p\in \cP} \{ \mathbb{E}_p [u(f)] - \mathbb{E}_p [u(g)] \} < 0,
\end{equation}
the completeness of $\not\succsim$ requires that the above strict inequality always holds.
This condition makes it impossible for any pair of acts $f$ and $g$ to satisfy $\max_{\cP \in \bP} \min_{p\in \cP} \{ \mathbb{E}_p [u(f)] - \mathbb{E}_p [u(g)] \} = 0$, that is, $f \sim g$.}
Moreover, this argument implies that the completeness of $\not\succsim$ is incompatible with \textit{reflexivity}.
To avoid such technical issues, we introduce the subrelation $\succsim^+$ of $\succsim$ defined as follows: 
For all $f, g\in \F$, $f \succsim^+ g$ if for all $h,h'$, there exists  $\delta > 0 $ such that for all $\varepsilon \in (0, \delta)$, 
\begin{equation*}
    (1- \varepsilon )f +  \varepsilon  h \succsim  (1- \varepsilon ) g + \varepsilon  h'.
\end{equation*}
According to this definition, the comparison between $f$ and $g$ is considered robust in the sense that, no matter which acts $h$ and $h'$ are used to add noise, the comparison remains unchanged for sufficiently small noise.
Using this relation, we first consider the negative counterpart of \textit{completeness}.

\begin{axiom}[Negative Completeness]
    For all $f, g\in \F$, $f \nrsuccsim  g$ or $g \nrsuccsim  f$. 
\end{axiom}

The following result shows that if we impose \textit{negative completeness} on generalized Bewley preferences, then the conjunctive generalized Bewley preferences can be characterized---that is, \textit{negative completeness} plays the same role as \textit{constant-bound transitivity}. 

\begin{proposition}
\label{pro:int_neg}
    A  binary relation $\succsim$ is a generalized Bewley preference and satisfies \textit{negative completeness} if and only if it is a conjunctive generalized Bewley preference. 
\end{proposition}

We next consider a negative counterpart of \textit{constant-bound transitivity}.
Since the issues from adopting $\not\succsim$ do not occur in this axiom, we use it directly.

\begin{axiom}[Negative Constant-Bound Transitivity]
    For all $f\in \F$ and $x,y\in X$, if $x\not\succsim f$ and $f\not\succsim y$, then $x\not\succsim y$. 
\end{axiom}

The generalized Bewley preferences that satisfy \textit{negative constant-bound transitivity} become disjunctive generalized Bewley preferences.

\begin{proposition}
\label{pro:uni_neg}
    A  binary relation $\succsim$ is a generalized Bewley preference and satisfies \textit{negative constant-bound transitivity} if and only if it is a disjunctive generalized Bewley preference. 
\end{proposition}

The proof of Proposition \ref{pro:uni_neg} follows from the direct equivalence between \textit{completeness} and \textit{negative constant-bound transitivity}, as presented below.

\begin{lem}
\label{lem:neg_tran}
    Suppose that $\succsim$ satisfies \textit{reflexivity}, \textit{unambiguous transitivity}, and \textit{independence}. 
    Then, $\succsim$ satisfies \textit{completeness} if and only if it satisfies \textit{negative constant-bound transitivity}. 
\end{lem}
By combining this result with Theorem \ref{thm:uni}, it follows that \textit{constant-bound transitivity} yields the disjunctive generalized Bewley preferences.

We can understand these results using the same geometric arguments as in Theorems \ref{thm:uni} and \ref{thm:int}.
Recall that under a generalized Bewley preference $\succsim$, $f\succsim g$ if and only if $u(f)-u(g)\in C$, where $C$ is a nonempty closed cone. 
By the definition of $\succsim^+$, we can find that $f\nrsuccsim g$ if and only if $u(f)-u(g)$ is in the complement of $C$ or its boundary.
Then, by applying the same geometric argument for \textit{completeness}, it follows that \textit{negative completeness} requires the complement of the complement of $C$---that is, $C$ itself---to be diagonal-slice convex. Hence, applying the argument of Theorem \ref{thm:int}, this axiom yields the conjunctive generalized Bewley model.
Similarly, \textit{negative constant-bound transitivity} requires the complement of $C$ to be diagonal-slice convex, yielding the disjunctive generalized Bewley model as in Theorem \ref{thm:uni}.


\section{Literature}
\label{sec:liter}

Preferences that may violate completeness and transitivity have been considered in the literature. 
\citet{aumann1962utility} argued that the assumption of completeness is too strong to describe economic behavior and examined risk preferences that may violate completeness while satisfying the other axioms in the expected utility theorem. 
Aumann showed that for such a preference, there exists a vNM function compatible in a weak sense. 
\citet{dubra2004expected} refined this result by showing that these preferences can be represented through the unanimity rule among multiple vNM functions. 
Risk preferences that may violate completeness and transitivity were examined by \citet{hara2019coalitional}. 
They showed that reflexive and independent preferences can be represented by the multi-utility version of generalized Bewley preferences.
\citet{evren2011multi} and \citet{nishimura2016utility} established similar results in a more general setup, supporting the strong connection between violations of rationality and the multiplicity of utility functions.

In the literature on decision-making under ambiguity, \citet{bewley2002knightian} introduced the unanimity rule among multiple priors (see also \citet{gilboa2010objective}).
\citet{lehrer2011justifiable} studied the for-some version of multi-prior preferences and, furthermore, showed that a class of preferences that may violate completeness and transitivity can be represented as the generalized Bewley preferences. 
Compared to them, this paper fills the gap among these models by examining the relationship between the degree of rationality and cognitive procedures. 
Beyond the above models, irrational preferences have been studied in many directions, including
variants of the unanimity rule (\citet{nascimento2011class,faro2015variational,hill2016incomplete}), 
multi-prior and multi-utility preferences (\citet{seidenfeld1995representation,nau2006shape,ok2012incomplete,galaabaatar2013subjective}), comparisons of the best and worst scenarios (\citet{echenique2022twofold,bardier2025hopingbestpreparingworst}). 

Utility functions parameterized by collections of sets of priors are also closely related to our paper (e.g., \citet{ke2020randomization,chandrasekher2022dual,akita2025random}).
Notably, \citet{nakamura2025cautiousdualselfexpectedutility} obtained a preference representation similar to the conjunctive generalized Bewley preferences. 
Their main axiom weakens \citeauthor{gilboa1989maxmin}'s (\citeyear{gilboa1989maxmin}) ambiguity aversion by requiring the ambiguity-averse attitude only when randomization makes a perfect hedge. 
Moreover, the geometric arguments underlying their model and the the conjunctive generalized Bewely preferences are common---the upper contour sets in both preferences can be represented as diagonal-slice convex cones. 
Compared to their paper, our theorems show that similar structures can be obtained from axioms about completeness and transitivity, which at first glance seems unrelated to the axiom of ambiguity aversion.


\section*{APPENDIX}
\renewcommand{\thesubsection}{A.\arabic{subsection}}
\setcounter{subsection}{0}

\subsection{Proof of Theorem \ref{thm:uni}}

\textit{Only-if part.}
Let $\succsim$ be a generalized Bewley preference that satisfies \textit{completeness}. 
Then, there exist a nonconstant affine function $u: X\to \bR$ and a belief collection $\mathbb{P}$ such that for all $f,g\in\F$, 
\begin{equation}
\label{eq:uni_maxmin}
    f\succsim g \iff \max_{\cP\in \bP} \min_{p \in \cP} \{ \mathbb{E}_p [u(f)] - \mathbb{E}_p [u(g)] \}  \geq 0. 
\end{equation}
It is sufficient to prove that for all $f, g\in \F$, 
\begin{equation}
\label{eq:uni_ints}
        \max_{\cP\in \bP} \min_{p \in \cP} \{ \mathbb{E}_p [u(f)] - \mathbb{E}_p [u(g)] \} \geq \min_{\cP\in \bP} \max_{p \in \cP} \{ \mathbb{E}_p [u(f)] - \mathbb{E}_p [u(g)] \} . 
    \end{equation}
    Indeed, if \eqref{eq:uni_ints} holds for all $f, g\in \F$, then 
    \begin{align*}
        & \max_{\cP\in \bP} \min_{p \in \cP} \{ \mathbb{E}_p [u(f)] - \mathbb{E}_p [u(g)] \} \\
        &=\max \qty{  \min_{\cP\in \bP} \max_{p \in \cP} \{ \mathbb{E}_p [u(f)] - \mathbb{E}_p [u(g)] \}  , ~ \max_{\cP\in \bP} \min_{p \in \cP} \{ \mathbb{E}_p [u(f)] - \mathbb{E}_p [u(g)] \} }. 
    \end{align*}
    Together with \eqref{eq:uni_maxmin}, we obtain the desired result. 
    
    Suppose to the contrary that there exist $f^\ast, g^\ast \in \F$ such that 
    \begin{equation}
    \label{eq:uni_contra1}
        \min_{\cP\in \bP} \max_{p \in \cP} \{ \mathbb{E}_p [u(f^\ast)] - \mathbb{E}_p [u(g^\ast)] \}  >  \max_{\cP\in \bP} \min_{p \in \cP} \{ \mathbb{E}_p [u(f^\ast)] - \mathbb{E}_p [u(g^\ast)] \}. 
    \end{equation}
    Without loss of generality, we can assume that 
    \begin{equation}
    \label{eq:uni_contra2}
        \min_{\cP\in \bP} \max_{p \in \cP} \{ \mathbb{E}_p [u(f^\ast)] - \mathbb{E}_p [u(g^\ast)] \}  > 0 >   \max_{\cP\in \bP} \min_{p \in \cP} \{ \mathbb{E}_p [u(f^\ast)] - \mathbb{E}_p [u(g^\ast)] \}. 
    \end{equation}
    By \eqref{eq:uni_maxmin}, $f^\ast \not\succsim g^\ast$. 
    On the other hand, by the first inequality of \eqref{eq:uni_contra2}, we have 
    \begin{equation}
        \label{eq:uni_contra3}
         0 >   \max_{\cP\in \bP} \min_{p \in \cP} \{ \mathbb{E}_p [u(g^\ast)] - \mathbb{E}_p [u(f^\ast)] \}. 
    \end{equation}
    Hence,  $g^\ast \not\succsim f^\ast$, which is a contradiction to \textit{completeness}. 

    \textit{If part.} Let $\succsim$ be a disjunctive generalized Bewley preference associated with $(u, \mathbb{P})$. We only verify \textit{completeness}. Take $f, g\in \F$ arbitrarily. It is sufficient to prove that if $f\not\succsim g$, then $g\succsim f$. 
    By the definition of disjunctive generalized Bewley preferences, 
    \begin{equation*}
         \max\qty{ \max_{\cP\in \bP} \min_{p \in \cP} \{ \mathbb{E}_p [u(f)] - \mathbb{E}_p [u(g)] \}, ~ \min_{\cP\in \bP} \max_{p \in \cP} \{ \mathbb{E}_p [u(f)] - \mathbb{E}_p [u(g)] \} } < 0. 
    \end{equation*}
    Then, we have 
    \begin{equation*}
         \min \qty{ \min_{\cP\in \bP} \max_{p \in \cP} \{ - \mathbb{E}_p [u(f)] + \mathbb{E}_p [u(g)] \}, ~ \max_{\cP\in \bP} \min_{p \in \cP} \{ - \mathbb{E}_p [u(f)] + \mathbb{E}_p [u(g)] \} } > 0. 
    \end{equation*}
    Thus, we have $g\succsim f$.

\subsection{Proof of Theorem \ref{thm:int}}

    \textit{Only-if part.}
    Let $\succsim$ be a generalized Bewley preference that satisfies \textit{constant-bound transitivity}. Then, there exist a nonconstant affine function $u: X\to \bR$ and a belief collection $\mathbb{P}$ such that for all $f,g\in\F$, 
    \begin{equation}
    \label{eq:ints_maxmin}
        f\succsim g \iff \max_{\cP\in \bP} \min_{p \in \cP} \{ \mathbb{E}_p [u(f)] - \mathbb{E}_p [u(g)] \}  \geq 0. 
    \end{equation}
    Without loss of generality, we assume that there exists $x_0\in X$ such that $u(x_0) = 0\in \text{int}\, u(X)$. 
    It is sufficient to prove that for all $f, g\in \F$, 
    \begin{equation}
    \label{eq:wts_ints}
        \min_{\cP\in \bP} \max_{p \in \cP} \{ \mathbb{E}_p [u(f)] - \mathbb{E}_p [u(g)] \}  \geq \max_{\cP\in \bP} \min_{p \in \cP} \{ \mathbb{E}_p [u(f)] - \mathbb{E}_p [u(g)] \}. 
    \end{equation}
    Indeed, if \eqref{eq:wts_ints} holds for all $f, g\in \F$, then 
    \begin{align*}
        & \max_{\cP\in \bP} \min_{p \in \cP} \{ \mathbb{E}_p [u(f)] - \mathbb{E}_p [u(g)] \} \\
        &=\min \qty{  \min_{\cP\in \bP} \max_{p \in \cP} \{ \mathbb{E}_p [u(f)] - \mathbb{E}_p [u(g)] \}  , ~ \max_{\cP\in \bP} \min_{p \in \cP} \{ \mathbb{E}_p [u(f)] - \mathbb{E}_p [u(g)] \} }. 
    \end{align*}
    Together with \eqref{eq:ints_maxmin}, we obtain the desired result. 

    Suppose to the contrary that there exist $f^\ast, g^\ast \in \F$ such that 
    \begin{equation}
    \label{eq:ints_contra1}
        \min_{\cP\in \bP} \max_{p \in \cP} \{ \mathbb{E}_p [u(f^\ast)] - \mathbb{E}_p [u(g^\ast)] \}  <  \max_{\cP\in \bP} \min_{p \in \cP} \{ \mathbb{E}_p [u(f^\ast)] - \mathbb{E}_p [u(g^\ast)] \}. 
    \end{equation}
    Without loss of generality, we can take $g^\ast$ such that $u(g^\ast)$ is a constant vector in $\mathbb{R}^S$. 
    Then, \eqref{eq:ints_contra1} can be rewritten as 
    \begin{equation}
    \label{eq:ints_contra2}
        \min_{\cP\in \bP} \max_{p \in \cP} \mathbb{E}_p [u(f^\ast)]   <  \max_{\cP\in \bP} \min_{p \in \cP} \mathbb{E}_p [u(f^\ast)]. 
    \end{equation}
    Let $f^{**} \in \F$ be such that for some $\beta \in \mathbb{R}_{++}$, 
    \begin{equation*}
        u(f^{**} (s)) = \beta (u(f^*(s))  -  \max_{\cP\in \bP} \min_{p \in \cP} \mathbb{E}_p [u(f^\ast)] ) \in u(X)
    \end{equation*}
    for each $s\in S$. 
    Then, 
    \begin{equation}
    \label{eq:ints_contra3}
        \min_{\cP\in \bP} \max_{p \in \cP} \mathbb{E}_p [u(f^{**})]   <  \max_{\cP\in \bP} \min_{p \in \cP} \mathbb{E}_p [u(f^{**})] = 0. 
    \end{equation}

    Let $\varepsilon \in \mathbb{R}$ be such that
    \begin{equation}
        0 < \varepsilon <  \max_{\cP\in \bP} \min_{p \in \cP} \mathbb{E}_p [u(f^{**})] -  \min_{\cP\in \bP} \max_{p \in \cP} \mathbb{E}_p [u(f^{**})].
    \end{equation}
    Define $x_{-\varepsilon} \in X$ as $u(x_{-\varepsilon}) = -\varepsilon$. 
    Then, we have 
    \begin{align*}
        \max_{\cP\in \bP} \min_{p \in \cP} \{ \mathbb{E}_p [u(x_{-\varepsilon})] - \mathbb{E}_p [u(f^{**})] \} 
        &= -\varepsilon -\min_{\cP\in \bP} \max_{p \in \cP} \mathbb{E}_p [u(f^{**})] \\
        &> -  \max_{\cP\in \bP} \min_{p \in \cP} \mathbb{E}_p [u(f^{**})] \\
        &= 0. 
    \end{align*}
    Therefore, we have $x_{-\varepsilon} \succsim f^{**}$. However, by \eqref{eq:ints_contra3} and the definitions of $x_0$ and $x_{-\varepsilon}$, we have $f^{**}\succsim x_0$ and $x_0 \succ x_{-\varepsilon}$. This is a contradiction to \textit{constant-bound transitivity}. 

    \textit{If part.} Let $\succsim$ be a conjunctive generalized Bewley preference associated with $(u, \mathbb{P})$. We only verify \textit{constant-bound transitivity}.
    Let $x,y\in X$ and $f\in \F$ such that $x\succsim f$ and $f\succsim y$. 
    By the definition, 
    \begin{align*}
        x\succsim f &\iff \min\qty{ \max_{\cP\in \bP} \min_{p \in \cP} \{ \mathbb{E}_p [u(x)] - \mathbb{E}_p [u(f)] \}, ~ \min_{\cP\in \bP} \max_{p \in \cP} \{ \mathbb{E}_p [u(x)] - \mathbb{E}_p [u(f)] \} } \geq 0\\
        &\iff u(x) - \max\qty{ \min_{\cP\in \bP} \max_{p \in \cP} \mathbb{E}_p [u(f)] , ~ \max_{\cP\in \bP} \min_{p \in \cP} \mathbb{E}_p [u(f)]  } \geq 0 \\
        &\iff u(x) \geq  \max\qty{ \max_{\cP\in \bP} \min_{p \in \cP} \mathbb{E}_p [u(f)] , ~ \min_{\cP\in \bP} \max_{p \in \cP} \mathbb{E}_p [u(f)]  }. 
    \end{align*}
    Similarly, 
    \begin{align*}
        f\succsim y &\iff \min \qty{ \max_{\cP\in \bP} \min_{p \in \cP} \{ \mathbb{E}_p [u(f)] - \mathbb{E}_p [u(y)] \}, ~ \min_{\cP\in \bP} \max_{p \in \cP} \{ \mathbb{E}_p [u(f)] - \mathbb{E}_p [u(y)] \} } \geq 0\\
        &\iff \min\qty{\max_{\cP\in \bP} \min_{p \in \cP}  \mathbb{E}_p [u(f)] , ~ \min_{\cP\in \bP} \max_{p \in \cP}  \mathbb{E}_p [u(f)]  }\geq u(y)
    \end{align*}
    Therefore, we have $u(x)\geq u(y)$, that is, $x\succsim y$.

\subsection{Proof of Theorem \ref{thm_half}}

\textit{Only-if part.} This part immediately follows from the proofs of Theorems \ref{thm:uni} and \ref{thm:int}. Let $\succsim$ be a generalized Bewley preference that satisfies \textit{completeness} and \textit{constant-bound transitivity}. Then, there exist a nonconstant affine function $u: X\to \bR$ and a belief collection $\mathbb{P}$ such that for all $f,g\in\F$, 
\begin{equation}
\label{eq:half_maxmin}
        f\succsim g \iff \max_{\cP\in \bP} \min_{p \in \cP} \{ \mathbb{E}_p [u(f)] - \mathbb{E}_p [u(g)] \}  \geq 0. 
\end{equation}
By the arguments in the proofs of Theorems \ref{thm:uni} and \ref{thm:int}, for all $f,g\in \F$, 
\begin{equation}
    \max_{\cP\in \bP} \min_{p \in \cP} \{ \mathbb{E}_p [u(f)] - \mathbb{E}_p [u(g)] \} 
    = 
    \min_{\cP\in \bP} \max_{p \in \cP} \{ \mathbb{E}_p [u(f)] - \mathbb{E}_p [u(g)] \}. 
\end{equation}
Then, for all $f,g\in \F$, 
\begin{align}
    &\max_{\cP\in \bP} \min_{p \in \cP} \{ \mathbb{E}_p [u(f)] - \mathbb{E}_p [u(g)] \} \\
    &= 
    {1\over 2}
    \max_{\cP\in \bP} \min_{p \in \cP} \{ \mathbb{E}_p [u(f)] - \mathbb{E}_p [u(g)] \} 
    + {1\over 2}
    \min_{\cP\in \bP} \max_{p \in \cP} \{ \mathbb{E}_p [u(f)] - \mathbb{E}_p [u(g)] \}. 
\end{align}
By \eqref{eq:half_maxmin}, we obtained the desired result.

\textit{If part.}
Let $\succsim$ be a half-mixture generalized Bewley preference associated with $(u, \bP)$. We prove that it satisfies \textit{completeness}. 
Let $f, g\in \mathcal{F}$. If $f \not \succsim g$, then 
\begin{equation*}
     {1\over 2} \max_{\cP\in \bP} \min_{p \in \cP} \{ \mathbb{E}_p [u(f)] - \mathbb{E}_p [u(g)] \} + {1\over 2} \min_{\cP\in \bP} \max_{p \in \cP} \{ \mathbb{E}_p [u(f)] - \mathbb{E}_p [u(g)] \} < 0. 
\end{equation*}
This is equivalent to 
\begin{equation*}
     {1\over 2} \min_{\cP\in \bP} \max_{p \in \cP} \{ \mathbb{E}_p [u(g)] - \mathbb{E}_p [u(f)] \} + {1\over 2} \max_{\cP\in \bP} \min_{p \in \cP} \{ \mathbb{E}_p [u(g)] - \mathbb{E}_p [u(f)] \} > 0, 
\end{equation*}
which implies $g\succsim f$.

Next, we prove that $\succsim$ satisfies \textit{constant-bound transitivity}. Let $f\in \mathcal{F}$ and $x, y\in X$ be such that $x \succsim f \succsim y$. 
By $x\succsim f$, we have 
\begin{align}
     & {1\over 2} \max_{\cP\in \bP} \min_{p \in \cP} \{ \mathbb{E}_p [u(x)] - \mathbb{E}_p [u(f)] \} + {1\over 2} \min_{\cP\in \bP} \max_{p \in \cP} \{ \mathbb{E}_p [u(x)] - \mathbb{E}_p [u(f)]\}\geq 0 \notag \\
     & \iff u(x) \geq -
     {1\over 2} \max_{\cP\in \bP} \min_{p \in \cP} \{  - \mathbb{E}_p [u(f)] \} - {1\over 2} \min_{\cP\in \bP} \max_{p \in \cP} \{  - \mathbb{E}_p [u(f)]\} \notag \\
     &\iff  u(x) \geq 
     {1\over 2} \min_{\cP\in \bP} \max_{p \in \cP} \{   \mathbb{E}_p [u(f)] \} + {1\over 2} \max_{\cP\in \bP} \min_{p \in \cP} \{  \mathbb{E}_p [u(f)]\}. \label{eq:half_xf}
\end{align}
On the other hand, $f\succsim y$, we have 
\begin{align}
     & {1\over 2} \max_{\cP\in \bP} \min_{p \in \cP} \{ \mathbb{E}_p [u(f)] - \mathbb{E}_p [u(y)] \} + {1\over 2} \min_{\cP\in \bP} \max_{p \in \cP} \{ \mathbb{E}_p [u(f)] - \mathbb{E}_p [u(y)]\}\geq 0 \notag \\
     &\iff  
     {1\over 2} \min_{\cP\in \bP} \max_{p \in \cP} \{   \mathbb{E}_p [u(f)] \} + {1\over 2} \max_{\cP\in \bP} \min_{p \in \cP} \{  \mathbb{E}_p [u(f)]\} \geq  u(y). \label{eq:half_fy}
\end{align}
By \eqref{eq:half_xf} and \eqref{eq:half_fy}, we have $u(x)\geq u(y)$, which means that $x\succsim y$. 

\subsection{Proof of Proposition \ref{pro:interchange}}

Let $\succsim$ be a generalized Bewley preference associated with $(u, \bP)$.
Suppose that $\succsim$ satisfies \textit{completeness} and \textit{constant-bound transitivity}. 
By the argument in the proofs of Theorems \ref{thm:uni} and \ref{thm:int}, for all $f, g\in \F$, 
\begin{equation}
\label{eq:commutative}
    \max_{\cP\in \bP} \min_{p \in \cP}\{  \mathbb{E}_p [u(f)] - \mathbb{E}_p [u(g)] \} =  \min_{\cP\in \bP} \max_{p \in \cP} \{  \mathbb{E}_p [u(f)] - \mathbb{E}_p [u(g)] \}. 
\end{equation}

If \eqref{eq:commutative} holds for all $f,g \in \F$, then
\begin{align*}
    &\max_{\cP\in \bP} \min_{p \in \cP}\{  \mathbb{E}_p [u(f)] - \mathbb{E}_p [u(g)] \} \\
    &= \max\qty{ \max_{\cP\in \bP} \min_{p \in \cP} \{ \mathbb{E}_p [u(f)] - \mathbb{E}_p [u(g)] \}, ~ \min_{\cP\in \bP} \max_{p \in \cP} \{ \mathbb{E}_p [u(f)] - \mathbb{E}_p [u(g)] \}  } \\
    &= \min\qty{ \max_{\cP\in \bP} \min_{p \in \cP} \{ \mathbb{E}_p [u(f)] - \mathbb{E}_p [u(g)] \}, ~ \min_{\cP\in \bP} \max_{p \in \cP} \{ \mathbb{E}_p [u(f)] - \mathbb{E}_p [u(g)] \}  }. 
\end{align*}
Therefore, $\succsim$ is the disjuctive generalized Bewley preference associated with $(u, \bP)$ and the conjunctive generalized Bewley preference associated with $(u, \bP)$. 
By Theorems \ref{thm:uni} and \ref{thm:int}, $\succsim$ satisfies \textit{completeness} and \textit{constant-bound transitivity}. 

\subsection{Proof of Proposition \ref{pro:seu}}

We only prove the sufficiency part since the necessity part is obviously satisfied.
Let $\succsim$ be a generalized Bewley preference associated with $(u, \bP)$.
Suppose that $\succsim$ satisfies \textit{completeness} and \textit{constant-bound transitivity}. 
Let $S=\{s_1,s_2\}$.
It suffices to show that there exists $p^\ast \in \Delta(S)$ such that $p^\ast(s_1)=\max_{\cP\in\bP}\min_{p\in\cP}p(s_1)=\min_{\cP\in\bP}\max_{p\in\cP}p(s_1)$.
Indeed, by Theorem \ref{thm_half}, for all $f,g\in\mathcal{F}$,
\begin{align*}
    f\succsim g &\iff {1\over 2}\max_{\cP\in \bP} \min_{p \in \cP} \{ \mathbb{E}_p [u(f)] - \mathbb{E}_p [u(g)] \} + {1\over 2} \min_{\cP\in \bP} \max_{p \in \cP} \{ \mathbb{E}_p [u(f)] - \mathbb{E}_p [u(g)] \}\geq 0 \\
    &\iff \frac{1}{2}\left\{u(f(s_1))-u(g(s_1))-u(f(s_2))+u(g(s_2))\right\}\max_{\cP\in\bP}\min_{p\in\cP}p(s_1) \\
    &~~~~~~~~~~
    + \frac{1}{2}\left\{u(f(s_1))-u(g(s_1))-u(f(s_2))+u(g(s_2))\right\}\min_{\cP\in\bP}\max_{p\in\cP}p(s_1) \\
    &~~~~~~~~~~
    +u(f(s_2))-u(g(s_2)) \geq 0 \\
    &\iff p^\ast (s_1)\left\{u(f(s_1))-u(g(s_1))-u(f(s_2))+u(g(s_2))\right\}+u(f(s_2))-u(g(s_2)) \geq 0 \\
    &\iff \mathbb{E}_{p^\ast}[u(f)] \geq \mathbb{E}_{p^\ast}[u(g)].
\end{align*}

Suppose to the contrary that $\max_{\cP\in\bP}\min_{p\in\cP}p(s_1) \neq \min_{\cP\in\bP}\max_{p\in\cP}p(s_1)$.
Without loss of generality, we assume that there exists $x_0\in X$ such that $u(x_0)=0\in \text{int}\:u(X)$.
Let $f\in\mathcal{F}$ be such that $u(f(s_1)) \neq 0$ and $u(f(s_2))=0$.
Then, 
\begin{align*}
    &\max_{\cP\in\bP}\min_{p\in\cP}p(s_1)u(f(s_1))\neq \min_{\cP\in\bP}\max_{p\in\cP}p(s_1)u(f(s_1)) \\
    &\iff \max_{\cP\in\bP}\min_{p\in\cP}\left\{\mathbb{E}_p [u(f)]-u(x_0)\right\} \neq \min_{\cP\in\bP}\max_{p\in\cP}\left\{\mathbb{E}_p [u(f)]-u(x_0)\right\}.
\end{align*}
This is a contradiction to Proposition \ref{pro:interchange}.

\subsection{Proof of Proposition \ref{prp:para_comp}}

Without loss of generality, we can assume that there exists $x_0 \in X$ such that $u(x_0) = 0 \in \text{int} ~ u(X)$. 
    
Suppose that $\succsim$ satisfies \textit{completeness} but $\mathbb{P}$ has a cutting hyperplane. 
Then, there exists a pair $(\varphi, \lambda) \in \mathbb{R}^S \times \mathbb{R}$ such that for each $\cP\in \mathbb{P}$, there exist $p^+,p^- \in \cP$ such that
\begin{equation}
\label{eq:cutting}
    \sum_{s\in S} p^+ (s) [ \varphi(s) - \lambda ] > 0 > \sum_{s\in S} p^- (s) [ \varphi(s) - \lambda ]. 
    \end{equation} 
Without loss of generality, we assume that $\varphi - \lambda \mathbf{1}_S \in u(X)^S$. 
Let $f\in \mathcal{F}$ be such that $u (f) = \varphi - \lambda \mathbf{1}_S$. 
By the first inequality of \eqref{eq:cutting}, 
\begin{equation}
    \min_{p\in \cP} \{ \mathbb{E}_p [u(x_0)] - \mathbb{E}_p [u(f)] \} < 0 ~~~~ \text{for all $\cP\in \mathbb{P}$}.
\end{equation}
Furthermore, by the second inequality of \eqref{eq:cutting}, 
\begin{equation}
    \min_{p\in \cP} \{ \mathbb{E}_p [u(f)] - \mathbb{E}_p [u(x_0)] \} < 0 ~~~~ \text{for all $\cP\in \mathbb{P}$}.
\end{equation}
Therefore, $f\not\succsim x_0$ and $x_0 \not\succsim f$, which contradicts \textit{completeness}. 

To prove the converse, suppose that $\mathbb{P}$ does not have a cutting hyperplane. Take $f, g\in \mathcal{F}$ arbitrarily. Then, there exists $\cP\in \mathbb{P}$ such that either (i) for all $p\in \cP$, 
$\sum_{s\in S} p (s) [  u(f(s)) - u(g(s)) ] > 0$ or (ii) for all $p\in \cP$, 
$\sum_{s\in S} p (s) [  u(f(s)) - u(g(s)) ] < 0$. 
In the case (i), we have $\max_{\cP\in \bP}\min_{p\in \cP} \{ \mathbb{E}_p [u(f)] - \mathbb{E}_p [u(g)] \} > 0$; and in the case (ii), we have $\max_{\cP\in \bP}\min_{p\in \cP} \{ \mathbb{E}_p [u(g)] - \mathbb{E}_p [u(f)] \} > 0$. 
Therefore, $f\succsim g$ or $g\succsim f$ holds. 

\subsection{Proof of Proposition \ref{prp:para_tran}}

We will use the following lemma by \citet{samet1998common}:


\begin{lem}
\label{lem:samet}
    Let $\cP_1,\cP_2\in\mathcal{K}(\Delta(S))$.
    Then, $\cP_1 \cap \cP_2=\emptyset$ if and only if there exist $\phi_1,\phi_2\in\bR^S$ such that $\phi_1+\phi_2=0$ and $\min_{p\in \cP_i}\mathbb{E}_{p}[\phi_i]>0$ for each $i=1,2$.
\end{lem}
    
Without loss of generality, we can assume that there exists $x_0 \in X$ such that $u(x_0) = 0 \in \text{int} ~ u(X)$. 

Suppose that $\succsim$ satisfies \textit{constant-bound transitivity} but there exist $\cP_1,\cP_2\in\mathbb{P}$ such that $\cP_1 \cap \cP_2 =\emptyset$.
Then, by Lemma \ref{lem:samet}, there exist $\phi_1,\phi_2\in\bR^S$ such that $\phi_1+\phi_2=0$, $\min_{p\in\cP_1}\mathbb{E}_p[\phi_1]>0$ and $\min_{p\in\cP_2}\mathbb{E}_{p}[\phi_2]>0$.
Without loss of generality, we assume $\phi_1\in u(X)^S$.
Let $f\in\mathcal{F}$ be an act such that $u(f)=\phi_1$ and $x_\varepsilon\in X$ be a lottery such that for $\varepsilon>0$ small enough, $u(x_\varepsilon)=\varepsilon$.
Then, 
\begin{equation}
    \max_{\cP\in \bP} \min_{p \in \cP} \{ \mathbb{E}_p [u(f)] - u(x_{\varepsilon}) \}  \geq \min_{p \in \cP_1} \{ \mathbb{E}_p [u(f)] - u(x_{\varepsilon}) \} =\min_{p \in \cP_1}\mathbb{E}_p[\phi_1]-\varepsilon>0
\end{equation}
and 
\begin{equation}
\label{eq:para_tran}
    \max_{\cP\in \bP} \min_{p \in \cP} \{ u(x_0)-\mathbb{E}_p [u(f)] \}  \geq \min_{p \in \cP_2} \{ u(x_0) - \mathbb{E}_p [u(f)] \} =\min_{p \in \cP_2} \mathbb{E}_p[\phi_2] >0.
\end{equation}
The equality of \eqref{eq:para_tran} holds since $\min_{p \in \cP_2} \{ u(x_0) - \mathbb{E}_p [u(f)] \}=\min_{p \in \cP_2} \mathbb{E}_p [-u(f)]=\mathbb{E}_p [-\phi_1]$.
Therefore, $x_0\succsim f\succsim x_\varepsilon$ holds.
\textit{Constant-bound transitivity} implies $x_0\succsim x_\varepsilon$, which contradicts $u(x_\varepsilon)>u(x_0)$.

To prove the converse, we consider the contrapositive.
Suppose that $\succsim$ does not satisfy \textit{constant-bound transitivity}. 
Then, there exist $x,y\in X$ and $f\in\mathcal{F}$ such that $x\succsim f \succsim y$ but $y\succ x$.
Since $u$ is an affine function, there exists $z\in X$ such that $u(y)>u(z)>u(x)$.
Then, $x\succsim f$ implies that for some $\cP_1\in \bP$, $\min_{p\in \cP_1} \{ u(x) - \mathbb{E}_p [u(f)] \} \geq 0 $ holds, which implies $\min_{p\in \cP_1} \{ u(z) - \mathbb{E}_p [u(f)] \} >0$. 
Furthermore, $f\succsim y$ implies that for some $\cP_2 \in \bP$,  $\min_{p\in \cP_2} \{ \mathbb{E}_p [u(f)] - u(y) \} \geq 0 $ holds, which implies $\min_{p\in \cP_2} \{ \mathbb{E}_p [u(f)] - u(z) \} >0$. 
Let $\phi_1,\phi_2\in\bR^S$ be such that $\phi_1=u(z)-u(f)$ and $\phi_2=u(f)-u(z)$.
Since we have $\phi_1+\phi_2=0$, $\min_{p\in\cP_1}\mathbb{E}_p[\phi_1]>0$, and $\min_{p\in\cP_2}\mathbb{E}_p[\phi_2]>0$, Lemma \ref{lem:samet} implies $\cP_1\cap \cP_2=\emptyset$.

\subsection{Proof of Proposition \ref{pro:int_neg}}

\textit{Only-if part.}
Let $\succsim$  be a generalized Bewley preference that satisfies \textit{negative completeness}. Then, there exist a nonconstant affine function $u: X\to \bR$ and a belief collection $\mathbb{P}$ such that for all $f,g\in\F$, 
\begin{equation}
\label{eq:nuni_maxmin}
    f\succsim g \iff \max_{\cP\in \bP} \min_{p \in \cP} \{ \mathbb{E}_p [u(f)] - \mathbb{E}_p [u(g)] \}  \geq 0. 
\end{equation}
As in the proof of Theorem \ref{thm:int}, it is sufficient to prove that for all $f, g\in \F$, 
\begin{equation}
\label{eq:nuni_ints}
    \min_{\cP\in \bP} \max_{p \in \cP} \{ \mathbb{E}_p [u(f)] - \mathbb{E}_p [u(g)] \} \geq \max_{\cP\in \bP} \min_{p \in \cP} \{ \mathbb{E}_p [u(f)] - \mathbb{E}_p [u(g)] \}. 
\end{equation}
    
Suppose to the contrary that there exist $f^\ast, g^\ast \in \F$ such that 
\begin{equation}
\label{eq:nuni_contra1}
    \max_{\cP\in \bP} \min_{p \in \cP} \{ \mathbb{E}_p [u(f^\ast)] - \mathbb{E}_p [u(g^\ast)] \} > \min_{\cP\in \bP} \max_{p \in \cP} \{ \mathbb{E}_p [u(f^\ast)] - \mathbb{E}_p [u(g^\ast)] \} . 
\end{equation}
Without loss of generality, we can assume that 
\begin{equation}
\label{eq:nuni_contra2}
    \max_{\cP\in \bP} \min_{p \in \cP} \{ \mathbb{E}_p [u(f^\ast)] - \mathbb{E}_p [u(g^\ast)] \} 
    > 0 > 
    \min_{\cP\in \bP} \max_{p \in \cP} \{ \mathbb{E}_p [u(f^\ast)] - \mathbb{E}_p [u(g^\ast)] \} . 
\end{equation}
By \eqref{eq:nuni_maxmin}, $f^\ast \nrsuccsim g^\ast$ does not hold. On the other hand, by \eqref{eq:nuni_contra2}, we have 
\begin{equation}
\label{eq:nuni_contra3}
    \max_{\cP\in \bP} \min_{p \in \cP} \{ \mathbb{E}_p [u(g^\ast)] - \mathbb{E}_p [u(f^\ast)] \} > 0 > \min_{\cP\in \bP} \max_{p \in \cP} \{ \mathbb{E}_p [u(g^\ast)] - \mathbb{E}_p [u(f^\ast)] \} . 
\end{equation}
Hence,  $g^\ast \nrsuccsim f^\ast$ does not hold, which is a contradiction to \textit{negative completeness}. 

\textit{If part.} Let $\succsim$ be a disjunctive generalized Bewley preference $(u, \mathbb{P})$. We only verify \textit{negative completeness}. Take $f, g\in \F$ arbitrarily. It is sufficient to prove that if $f\nrsuccsim g$ does not hold (i.e., $f\succsim^+ g$), then $g \nrsuccsim f$. 
By the definition of conjunctive generalized Bewley preferences, 
\begin{equation*}
    \min\qty{ \max_{\cP\in \bP} \min_{p \in \cP} \{ \mathbb{E}_p [u(f)] - \mathbb{E}_p [u(g)] \}, ~ \min_{\cP\in \bP} \max_{p \in \cP} \{ \mathbb{E}_p [u(f)] - \mathbb{E}_p [u(g)] \} } >  0. 
\end{equation*}
Then, we have 
\begin{equation*}
    \max \qty{ \min_{\cP\in \bP} \max_{p \in \cP} \{ - \mathbb{E}_p [u(f)] + \mathbb{E}_p [u(g)] \}, ~ \max_{\cP\in \bP} \min_{p \in \cP} \{ - \mathbb{E}_p [u(f)] + \mathbb{E}_p [u(g)] \} } <  0. 
\end{equation*}
Thus, we have $g \nrsuccsim f$. 

\subsection{Proof of Lemma \ref{lem:neg_tran}}

Suppose that $\succsim$ satisfies \textit{completeness}. 
If $\succsim$ violates \textit{negative constant-bound transitivity}, then $x\not\succsim f \not\succsim y$ but $x\succsim y$ for some $x,y\in X$ and some $f\in \F$. 
By \textit{completeness}, $y\succ f\succ x$. 
By \textit{unambiguous transitivity}, $x\succsim f$ holds, which is a contradiction to $x\not\succsim f$. 
    
To prove the converse, suppose that $\succsim$ satisfies \textit{negative constant-bound transitivity}. 
If $\succsim$ violates \textit{completeness}, then there exist $f,g\in \F$ such that $f\not\succsim g$ and $g\not\succsim f$. 
Let $x\in X$ be such that there exists $h\in \F$ with $x \sim  {1\over 2} g(s) + {1\over 2} h(s)$ for all $s\in S$. 
By \textit{independence}, ${1\over 2}f + {1\over 2} h\not\succsim {1\over 2}g + {1\over 2} h$ and ${1\over 2}g + {1\over 2} h\not\succsim {1\over 2}f + {1\over 2} h$ hold.
If $x \succsim {1\over 2}f + {1\over 2} h$, then \textit{unambiguous transitivity} implies ${1\over 2}g + {1\over 2} h \succsim{1\over 2}f + {1\over 2} h$, which is a contradiction. Similarly, ${1\over 2}f + {1\over 2} h \succsim x$ yields a contradiction. 
Therefore, we have ${1\over 2}f + {1\over 2} h\not\succsim x$ and $x\not\succsim {1\over 2}f + {1\over 2} h$. 
By \textit{negative constant-bound transitivity}, we have $x \not\succsim x$, which is a contradiction to \textit{reflexivity}. 

\bibliographystyle{econ}
\bibliography{reference}

\end{document}